\documentclass[11pt,twoside]{article}
\usepackage{float}
\usepackage[colorlinks,citecolor=blue,urlcolor=blue,linkcolor=blue,bookmarks=false]{hyperref}
\usepackage{amsfonts,epsfig,graphicx}
\usepackage{afterpage}
\usepackage{comment}
\usepackage{amsmath,amssymb,amsthm} 
\usepackage{bm}
\usepackage{fullpage}
\usepackage[T1]{fontenc} 
\usepackage{epsf} 
\usepackage{graphics} 
\usepackage{amsfonts,amsmath}
\usepackage[sort]{natbib} 
\usepackage{psfrag,xspace}
\usepackage{color,etoolbox}
\usepackage{algorithm}
\usepackage{algorithmic}
\usepackage{enumitem}
\usepackage{authblk}
\usepackage[capitalise,noabbrev]{cleveref}
\usepackage{graphicx}
\usepackage{subcaption}

\newcommand{\E}{\mathbb{E}}

\newcommand{\PP}{\mathbb{P}}
\newcommand{\QQ}{\mathbb{Q}}

\newcommand{\bmZ}{\bm{Z}}

\newcommand{\Yabar}{Y^{\overline{a}}}
\newcommand{\abar}{\overline{a}}
\newcommand{\Xbar}{\overline{X}}
\newcommand{\Abar}{\overline{A}}

\newcommand{\fipw}{f^{\abar}_{\text{IPW}}}

\def\logit{\text{logit}}
\def\expit{\text{expit}}

\DeclareSymbolFont{bbold}{U}{bbold}{m}{n}
\DeclareSymbolFontAlphabet{\mathbbold}{bbold}

\newtheorem{theorem}{Theorem}
\newtheorem{lemma}{Lemma}

\newtheorem{proposition}{Proposition}
\theoremstyle{definition}
\newtheorem{definition}{Definition}

\theoremstyle{remark}

\newtheorem{remark}{Remark}

\newtheorem*{assumption*}{\assumptionnumber}
\providecommand{\assumptionnumber}{}
\makeatletter
\newenvironment{assumption}[2]
 {%
  \renewcommand{\assumptionnumber}{Assumption $\mathcal{#1}$#2}%
  \begin{assumption*}%
  \protected@edef\@currentlabel{$\mathcal{#1}$#2}%
 }
 {%
  \end{assumption*}
 }
\makeatother

\begin{document}

\def\spacingset#1{\renewcommand{\baselinestretch}%
{#1}\small\normalsize} \spacingset{1}

\raggedbottom
\allowdisplaybreaks[1]

%%%%%%%%%%%%%%%%%%%%%%%%%%%%%%%%%%%%%%%%%%%

\title{\vspace*{-.4in} {Sensitivity analysis for incremental effects, with \\
application to a study of victimization \& offending}}
\author{
    Shuying Shen\thanks{Department of Statistics \& Data Science, Carnegie Mellon University.}, \ %\texttt{shuyings@andrew.cmu.edu}, \texttt{edward@stat.cmu.edu}.},
    Valerio Bacak\thanks{School of Criminal Justice, Rutgers University.}, \ 
    Edward H.\ Kennedy\protect\footnotemark[1] % \texttt{valerio.bacak@rutgers.edu}}
    }
\date{}

  \maketitle
  \thispagestyle{empty}

\begin{abstract}
Sensitivity analysis for unmeasured confounding under incremental propensity score interventions remains relatively underdeveloped. Incremental interventions define stochastic treatment regimes by multiplying the odds of treatment, offering a flexible framework for causal effect estimation. To study incremental effects when there are unobserved confounders, we adopt Rosenbaum's sensitivity model in single time point settings, and propose a doubly robust estimator for the resulting effect bounds. The bound estimators are asymptotically normal under mild conditions on  nuisance function estimation. We show that incremental effect bounds can be narrower or wider than those for mean potential outcomes, and that the bounds must lie between the expected minimum and maximum of the conditional bounds on $\E(Y^0\mid X)$ and $\E(Y^1\mid X)$. For time‐varying treatments, we consider the marginal sensitivity model. Although sharp bounds for incremental effects are identifiable from longitudinal data under this model, practical estimators have not yet been established; we discuss this challenge and provide partial results toward implementation. Finally, we apply our methods to study the effect of victimization on subsequent offending using data from the National Longitudinal Study of Adolescent to Adult Health (Add Health), illustrating the robustness of our findings in an empirical setting.
\end{abstract}


\section{Introduction}
The assumption of no unmeasured confounding is central to causal inference; it requires that, conditional on observed covariates, treatment assignment is independent of the potential outcomes. In practice, however, this assumption is often violated. Relevant confounders may be unmeasured or inherently unmeasurable. A substantial methodological literature has developed to address these challenges by relaxing the no unmeasured confounding assumption. 
We refer to \citet{nabi2024semiparametric} for a literature review; as discussed there, one can broadly separate into (a) point identification approaches (where confounding is sufficiently parameterized so that effects are point identified, depending on sensitivity parameter values, e.g., \citet{imbens2003sensitivity, robins2000sensitivity}), and (b) set identification approaches (without sensitivity parameters, e.g., \citet{robins1989analysis, manski1990nonparametric}, or with sensitivity parameters, e.g., \citet{rosenbaum1987sensitivity, yadlowsky2022bounds}) where causal effects are only bounded, not point identified. See \citet{nabi2024semiparametric} and references therein for many other examples of related and different approaches. Our approach falls within the second stream, of set identification with sensitivity parameters; we use a relaxation of the no unmeasured confounding assumption that depends on a sensitivity parameter, and bounds are constructed across a range of parameter values. \\

In particular, two sensitivity models play a crucial role in our work. The first is  Rosenbaum's sensitivity analysis model, which bounds the odds of treatment under different levels of unmeasured confounders \citep{rosenbaum2002observational}. This model is especially well suited to incremental propensity score interventions, which perturb the odds of treatment to assess how causal effects change under controlled deviations from the observed assignment mechanism. Accordingly, bounding the treatment odds ratio provides a natural framework for sensitivity analysis in this setting. A substantial literature has developed around this model. For example, \citet{yadlowsky2022bounds} develop a semi-parametric method to estimate  bounds on the conditional and average treatment effects, which achieve root-n rates if the nuisance functions are estimated at $n^{-1/4}$ rates. The second model of interest is the marginal sensitivity model, which can be equivalently formulated as bounding the odds ratio of treatment assignment between units with different potential outcomes \citep{tan2006distributional}. \citet{dorn2023sharp} derive sharp bounds under the marginal sensitivity model, and the comparison between these two odds ratio sensitivity model is discussed in \citet{tan2024model}. \citet{zhang2025enhanced} propose an enhanced marginal sensitivity model with tightened bounds. \\

All of the above sensitivity analysis literature has focused on deterministic interventions. In contrast, the incremental propensity score intervention \citep{kennedy2019nonparametric} is a stochastic intervention that operates by multiplicatively shifting the odds of treatment by a factor $\delta\in(0, \infty)$ relative to the observed treatment assignment mechanism. This construction yields a clear and interpretable estimand, allowing causal effects to be evaluated across a continuum of intervention intensities. Moreover, identification of incremental effects does not require strict positivity; instead, only weak positivity conditions are needed. Despite these attractive properties, sensitivity analysis for incremental effects remains limited. One exception is \citet{levis2024stochastic}, who develop estimators for sharp bounds on incremental effects under the marginal sensitivity model in single time point settings.\\

In this paper, we develop bounds for incremental effects in the presence of unmeasured confounding under the two sensitivity models described above. For settings with a single time point, we adopt Rosenbaum’s sensitivity model to identify upper and lower bounds on incremental effects and propose a doubly robust estimator for these bounds. Under mild regularity conditions, we establish asymptotic normality of the proposed estimator. For settings with time‐varying treatments, we work within the marginal sensitivity model. We derive an equivalent representation of this model and use it to characterize sharp bounds for incremental effects. Although these bounds are identifiable from observed longitudinal data, construction of a feasible estimator presents technical challenges that we leave for future work. Finally, we present an empirical application examining the causal effect of victimization on subsequent offending behavior.\\

\section{Preliminaries}

In this section we introduce the causal framework and notation used throughout the paper. We first review standard assumptions in causal inference and define incremental propensity score interventions, highlighting their interpretation and weak positivity requirements. We then introduce Rosenbaum’s sensitivity model and summarize existing results on bounding counterfactual means under unmeasured confounding, which form the building blocks for our subsequent analysis. \\

Three central identifying assumptions are commonly often made in causal inference: consistency, no unmeasured confounding (exchangeability), and positivity. \\

\begin{assumption}{A}{1}\label{A1}{\normalfont (\textbf{Consistency})}
$$
Y=Y^a \text{ if } A=a.
$$
\end{assumption}
\begin{assumption}{A}{2}\label{A2}{\normalfont (\textbf{No unmeasured confounding})}
$$
A\perp\!\!\!\perp Y^a\mid X.
$$
\end{assumption}
\begin{assumption}{A}{3}\label{A3}{\normalfont (\textbf{Positivity})}
$$
P(A=a \mid X) > 0.
$$
\end{assumption}

\bigskip

The consistency assumption implies no interference between individuals; the outcome of one individual is not influenced by the treatment of another individual. No unmeasured confounding requires that all relevant confounders of the treatment-outcome relationship are measured. Positivity means that every individual has a non-negative chance of receiving any level of treatment. 
In this paper, we avoid positivity by studying  incremental effects, and we avoid no unmeasured confounding by focusing on sensitivity analysis. Thus we really only rely on consistency, along with a relaxed version of no unmeasured confounding, to be described shortly. \\

To explore the effect of treatment under natural shifts of the treatment distribution, we use incremental propensity score interventions \citep{kennedy2019nonparametric}. These interventions are based on multiplying the odds of treatment by some factor $\delta\in(0, \infty)$. They can be conceptualized as setting the treatment to be a draw $Q(\delta)$ from the tilted conditional distribution $Q(\delta) \mid X \sim \text{Bern}\{ q(X;\delta,\pi)\} $, where the propensity score for the incremental intervention is
\begin{equation*}
    q(X;\delta,\pi)=\frac{\delta \pi(X)}{\delta\pi(X)+1-\pi(X)}.
\end{equation*}
Crucially, $\delta$ is an odds ratio, indicating how the intervention changes the odds of treatment, since
\begin{equation*}
    \delta = \frac{q(X)}{1-q(X)}\Big/\frac{\pi(X)}{1-\pi(X)}.
\end{equation*}
The range for the value of $\delta\in (0,\infty)$ defines the estimand, and depends on domain knowledge, e.g., based on what intensity of shifts are possible or of interest. For $\delta<1$, the intervention shifts the distribution of the propensity score downwards. For $\delta=1$, the treatment process is unchanged. For $\delta>1$, the odds of being treated increase. 
Importantly, the incremental propensity score intervention does not require positivity,  
since those with $\pi\in\{0,1\}$ are not intervened upon. \\

In the following definition we introduce Rosenbaum's sensitivity model \citep{rosenbaum2002observational}.\\

\begin{definition}\label{def:Rosenbaum}
Rosenbaum's sensitivity model assumes that the distribution $P$ satisfies the $\Gamma$-selection bias condition with some $\Gamma\in[1,\infty)$, i.e., for almost all $u,\tilde{u},X$, 
\begin{align}\label{eq:gammacon}
    \frac{1}{\Gamma} \leq \frac{P(A = 1 \mid X, U=u)}{P(A = 0 \mid X,U=u)}
    \frac{P(A = 0 \mid  X,U=\tilde{u})}{P(A = 1 \mid X, U=\tilde{u})}\leq \Gamma,
\end{align}
where $U$ are unmeasured confounders satisfying
$A\perp\!\!\!\perp Y^a\mid X, U$. 
\end{definition}

\bigskip

The above model therefore bounds the odds ratio of treatment for all possible values of unmeasured confounders; it is by now a popular and widely used model for unmeasured confounding. \\

According to the consistency assumption \ref{A1}, we define $\mu_a(X):=\E[Y\mid A=a,X]=\E[Y^a\mid A=a,X]$, which can be directly estimated from the data. However, due to the lack of exchangeability, we cannot identify $\E[Y^1\mid A=0,X]$ or $\E[Y^0\mid A=1,X]$ without further structure. \citet{yadlowsky2022bounds} characterize the lower and upper bounds for $\E[Y^1\mid A=0,X]$ and $\E[Y^0\mid A=1,X]$, denoted by $\theta_1^\pm(X)$  and $\theta_0^\pm(X)$, respectively. They show that the lower bound $\theta_1^-(X)$ for $\E[Y^1\mid A=0,X]$ can be  expressed as
\begin{align*}
  \theta_1^-(X) =
  \inf \left\{ \E[Y(1) L(Y(1)) \mid A = 1, X] ~:~ L \in \mathcal{L}\right\}
\end{align*}
where
\begin{align*}
   \mathcal{L} = \left\{ L :\mathbb{R} \to \mathbb{R}~\text{measurable} ~:~\begin{aligned} & 0 \le L(y) \le \Gamma L(\tilde{y})~\text{for all}~y,\tilde{y},\\&\E[L(Y(1)) \mid A=1, X] = 1\end{aligned}\right\}.
\end{align*}
The lower bound $\theta_1^-(X)$ for $\E[Y^1\mid A=0,X]$ can equivalently be expressed as the solution to the minimization problem
\begin{align*}
    \min_{\theta} \E\left[(Y^1-\theta)_+^2 +\Gamma (Y^1-\theta)_-^2 \mid A=1 \right],
\end{align*}
or the unique solution to 
\begin{align}\label{eq:cond0}
    \E\left[ f_{\theta_1^-}(Y^1) \mid A=1, X\right] = 0
\end{align}
if the solution is unique, where $f_{\theta}(y) = (y-\theta)_+-\Gamma(y-\theta)_-$. Similar corresponding expressions can be derived for $\theta_1^+(X)$ and $\theta_0^\pm(X)$. \\

\section{Sensitivity model}

In this section we develop identification results for incremental effects under unmeasured confounding. Focusing on single time point settings, we derive upper and lower bounds for incremental effects under Rosenbaum’s sensitivity model and show that these bounds are identifiable from observed data. We then compare the resulting bounds with those for mean potential outcomes,  establishing that incremental effect bounds can be narrower or wider, depending on the underlying distribution. \\

\subsection{Identification of bounds}
By definition of incremental interventions, the incremental effect can be expressed as
\begin{equation}\label{eq:id}
\begin{aligned}
    \psi(\delta)&=\E\left[\mathbb{E}\left[Y^{Q(\delta)}\mid X\right]\right]\\
    &=\E\left[\E\left[Y^1Q(\delta)+Y^0(1-Q(\delta))\mid X\right]\right]\\
    &=\E\left[q(X;\delta)\E[Y^1\mid X] + 
    (1-q(X;\delta)) \E[Y^0\mid X]\right].
\end{aligned}
\end{equation}
For identification of the desired parameter in \eqref{eq:id}, $\E[Y^0\mid X]$ and $\E[Y^1\mid X]$ need to be identified. Note that 
\begin{equation}\label{eq:mu}
\begin{aligned}
    \E[Y^1\mid X]&=\pi(X)\E[Y^1\mid A=1,X]+(1-\pi(X))\E[Y^1\mid A=0,X],\\
    \E[Y^0\mid X]&=(1-\pi(X))\E[Y^0\mid A=0,X]+\pi(X)\E[Y^0\mid A=1,X].
\end{aligned}
\end{equation}
The following proposition gives identifiable bounds for the incremental effect under only the consistency assumption \ref{A1}. \\

\begin{proposition}\label{prop:id1}
Under Assumption~\ref{A1}, lower and upper bounds $\psi^\pm(\delta)$ with $\psi^-(\delta)\leq\psi(\delta)\leq\psi^+(\delta)$ are given by
\begin{equation*}
\begin{aligned}
    \psi^\pm(\delta) 
    =&\mathbb{E}\Big[q(X)\pi(X)\mu_1(X)+q(X)(1-\pi(X))\theta_1^\pm(X) \\
    &+(1-q(X))(1-\pi(X))\mu_0(X)+(1-q(X))\pi(X)\theta_0^\pm(X)\Big].
\end{aligned}
\end{equation*}
\end{proposition}

The above proposition is a direct result from \eqref{eq:id} and \eqref{eq:mu}. This might not be a tight bound, because it combines  bounds on $\E[Y^1\mid A=0,X]$ and $\E[Y^0\mid A=1,X]$, which in general are only tight if the potential outcomes are symmetric, i.e., $Y^0 \overset{d}{=} C(1 - Y^1)$ \citep{yadlowsky2022bounds}. \\

\subsection{Length of the bounds}

Here we compare the bound length for incremental effects to those of mean potential outcomes. Although incremental effects can be viewed as arising under softer interventions closer to the observational one, somewhat surprisingly, there is no clear ordering of the bound lengths; incremental effects can yield wider or narrower bounds for finite non-zero $\delta$. \\

The bound length for incremental effects is
\begin{align*}
    L_{incremental} &= \E \Big[q(X)(1-\pi(X))(\theta_1^+(X)-\theta_1^-(X)) + (1-q(X))\pi(X)(\theta_0^+(X)-\theta_0^-(X))\Big] \\
    &= \E\left[\frac{\pi(X)(1-\pi(X))}{\delta\pi(X)+1-\pi(X)}\big(\delta(\theta_1^+(X) - \theta_1^-(X)) + (\theta_0^+(X) - \theta_0^-(X))\big)\right],
\end{align*}
and the bound lengths for mean potential outcomes are
\begin{align*}
    L_1 &:= L_{\E Y^1} = \E[(1-\pi(X))(\theta_1^+(X)-\theta_1^-(X))], \\
    L_0 &:= L_{\E Y^0} = \E[\pi(X)(\theta_0^+(X)-\theta_0^-(X))].
\end{align*}

As $\Gamma$ increases, $\theta_a^+ - \theta_a^-$ increases as well, and the bounds get wider. Note also that as $\delta \rightarrow \infty$ we have $q \rightarrow 1$, and $L_{incremental} \rightarrow L_1$, so the bounds align. The same occurs with $L_0$ as $\delta \rightarrow 0$ and $q \rightarrow 0$. The differences in the bound lengths are given by
\begin{align*}
    L_1 - L_{incremental} &= \E\Big[(1-q(X))\big[(1-\pi(X))(\theta_1^+(X)-\theta_1^-(X)) - \pi(X)(\theta_0^+(X)-\theta_0^-(X))\big]\Big], \\
    L_0 - L_{incremental} &= -\E\Big[q(X)\big[(1-\pi(X))(\theta_1^+(X)-\theta_1^-(X)) - \pi(X)(\theta_0^+(X)-\theta_0^-(X))\big]\Big]
\end{align*}
respectively. Since there is no general ordering of $\theta_1^+(X)-\theta_1^-(X)$ versus $\theta_0^+(X)-\theta_0^-(X)$, one cannot say one bound will generally be wider or narrower. The intuition is that incremental effects are a mix of mean potential outcomes under treatment and control, and so uncertainty in the control part can make the bounds wider or narrower than those under just treatment, and vice versa. \\

The same conclusion can also be seen from studying the behavior of the bound length of the incremental effects in $\delta$. By the dominated convergence theorem, the derivative $\frac{\partial}{\partial\delta}L_{incremental}$ with respect to the increment parameter $\delta$ is given by
\begin{align}\label{eq:L'}
    \E\left[\frac{\pi(X)(1-\pi(X))}{(\delta\pi(X)+1-\pi(X))^2}\Big[\big(1-\pi(X)\big)\big(\theta_1^+(X)-\theta_1^-(X)\big)-\pi(X)\big(\theta_0^+(X)-\theta_0^-(X)\big)\Big]\right].
\end{align}
Note the sign of this derivative is not fixed in $\delta$, and so the length may not be monotonic in $\delta$. See Section~\ref{sec:sim} for different patterns of the bound length. \\

There is a simple relation we can state based on the fact that incremental effects are a weighted combination of mean potential outcomes under treatment and control. Specifically, according to Equation~\ref{eq:id} we have the mixture relation
\begin{align*}
    \E\left[Y^Q(\delta)\mid X\right] = q(X)\E[Y^1\mid X] + (1-q(X))\E[Y^0\mid X].
\end{align*}
Therefore the upper and lower bounds of the conditional incremental effects satisfy
\begin{align*}
    q(X) \mu_1^\pm(X) + (1-q(X))\mu_0^\pm(X) \in \left[ \min_a\mu_a^-(X), \max_a\mu_a^+(X)\right],
\end{align*}
where $\mu_a^\pm(X)$ denote the bounds on the conditional potential outcome $\E(Y^a\mid X)$, with
\begin{align*}
    \mu_1^\pm(X) &= \pi(X)\mu_1(X)+(1-\pi(X))\theta_1^\pm(X), \\
    \mu_0^\pm(X) &= (1-\pi(X))\mu_0(X) + \pi(X)\theta_0^\pm(X).
\end{align*}
In words, the conditional incremental effect bounds must lie between the minimum and maximum of the conditional mean potential outcome bounds, since they are a weighted combination of both. As a result we also have the marginal relationship
\begin{align*}
    \psi^\pm(\delta) \in \left[\E\left[\min_a\mu_a^-(X)\right], \E\left[\max_a\mu_a^+(X)\right] \right].
\end{align*}

\bigskip

\section{Estimation and inference}

In this section we present estimation and inferential methods for the proposed bounds. We first derive the efficient influence function for the incremental effect bounds and construct a doubly robust estimator that uses sample splitting. We then establish asymptotic normality under mild regularity conditions. \\

\subsection{Efficiency bounds}

A crucial fact from \citet{yadlowsky2022bounds} is that, in a discrete nonparametric model, the influence function for the conditional lower bound $\theta_a^-(x)$ is given by 
\begin{equation*}
    IF(\theta_a^-(x)) = \frac{1(A=a) 1(X=x)}{\mathbb{P}(X=x)\pi(x)}\cdot
    \frac{f_{\theta_a^-(x)}(Y)}{\nu^-_a(x)}
\end{equation*}
where $f_{\theta}(y) = (y-\theta)_+-\Gamma(y-\theta)_-$ and 
\begin{equation*}
    \nu^-_a(x) = \PP(Y\geq\theta^-_a(X)\mid A=a, X=x) + \Gamma\PP(Y<\theta^-_a(x)\mid A=a,X=x).
\end{equation*}
Replacing $\Gamma$ with $1/\Gamma$ in the definition of $f$ and $\nu$, the influence function for $\theta^+_a(X)$ is of the same form. The next result combines this fact using the strategies of \citet{kennedy2024semiparametric} to derive the influence function for the marginalized bounds. \\

\begin{lemma}\label{lemma:EIF}
Suppose Assumption~\ref{A1} holds. Then in a nonparametric model the uncentered influence function for $\psi^\pm(\delta)$ is given by 
\begin{equation*}%\label{eq:IF}
\begin{aligned}
    \varphi^\pm(Z) =&\frac{\delta\pi(AY+(1-A)\theta_1^\pm)+(1-\pi)((1-A)Y+A\theta_0^\pm)}{\delta\pi+1-\pi} \\
    & + \frac{\delta(1-\pi)A}{\delta\pi+1-\pi}\cdot\frac{f_{\theta_1^\pm}(Y)}{\nu_1^\pm} +\frac{\pi(1-A)}{\delta\pi+1-\pi}\cdot\frac{f_{\theta_0^\pm}(Y)}{\nu_0^\pm} \\
    & + \frac{\delta(A-\pi)}{(\delta\pi+1-\pi)^2}(\pi\mu_1+(1-\pi)\theta_1^\pm-(1-\pi)\mu_0-\pi\theta_0^\pm).
\end{aligned}
\end{equation*}
\end{lemma}

\bigskip

Importantly, this efficient influence function depends on several nuisance functions: the propensity score $\pi$, regression functions $\mu_a$, conditional  bounds $\theta_a^\pm$ and summed conditional probabilities $\nu_a^\pm$. It should also be noted that the definition of the nuisance $\nu_a^\pm$ depends on the quantity $\theta_a^\pm$, which makes estimation non-trivial.\\

\subsection{Estimator}

Using the efficient influence function derived in the previous subsection, here we propose a new doubly robust-style estimator of the incremental effect bounds $\psi^\pm(\delta)$. \\

\begin{definition}
Our proposed doubly robust estimator for $\psi^\pm(\delta)$ is defined as the sample average of the estimated uncentered influence function, i.e.,  
\begin{equation*}
\widehat{\psi}^\pm(\delta)=\frac{1}{K}\sum^K_{k=1}\PP^k_n\{\varphi^\pm(Z;\widehat{\eta}_{-k}, \delta)\}.
\end{equation*}
Here $\PP_n$ denotes the sample mean, i.e., $\PP_n(s)=\frac{1}{n}\sum_i s(Z_i)$ for general functions $s$, and $\PP^k_n$ denotes the sample mean over the $k$th fold. The nuisances are given by the vector $\eta=(\pi,\mu_1,\mu_0,\theta_1^\pm, \theta_0^\pm, \nu_1^\pm,\nu_0^\pm)$ and $\widehat\eta_{-k}$ denotes estimates constructed from all folds except fold $k$. 
\end{definition}

\bigskip

The procedure to estimate the bounds $\psi^\pm(\delta)$, including sample splitting and estimation of nuisance functions, is described below in more detail. \\

\begin{algorithm}[H]
\caption{Doubly-robust estimator of bounds $\psi^\pm(\delta)$}
\begin{algorithmic}[1]\label{algo}
    \STATE Randomly split the data into $K$ folds.
    \STATE For $k=1,\cdots,K$,
    \begin{enumerate}[label=(\alph*), leftmargin=2em]
        \item Estimate $\pi,\mu_1,\mu_0$ with the $-k$ folds (all but the $k$th fold).
        \item Split the $-k$ folds into two halves. Use the first half to estimate $\theta_1^\pm,\theta_0^\pm$. Then use the second half, along with the estimated functions $\hat{\theta}_1^\pm,\hat{\theta}_0^\pm$, to estimate $\nu_1^\pm,\nu_0^\pm$.
        \item Compute the doubly robust estimator in the $k$th fold, $\PP^k_n\{\varphi^\pm(Z;\widehat{\eta}_{-k}, \delta)\}$.
    \end{enumerate}
    \STATE Average the $K$ estimators to obtain $\widehat\psi^\pm(\delta)$.
\end{algorithmic}
\end{algorithm}

\bigskip

$K=10$ folds is most commonly used in practice.  There are numerous possible methods to estimate the nuisance functions, including logistic regression, kernel/series methods, random forests, lasso, neural networks, ensembles/aggregates of the above, etc. \\

\citet{levis2024stochastic} also recently proposed an estimator for bounds on the incremental effect. However, their sensitivity model differs from ours in \eqref{eq:gammacon}. Specifically, they propose sharp bounds under the marginal sensitivity model
$$
    \frac{1}{\Gamma} \leq 
    \frac{\PP(A=1\mid X) / \PP(A=0\mid X)}{\PP(A=1\mid X, Y) / \PP(A=0\mid X, Y)}
    \leq \Gamma
$$
from \citet{zhao2019sensitivity}. 
We  consider a similar marginal sensitivity model generalized to longitudinal data in Section~\ref{sec:longitudinal}. \\

\subsection{Asymptotic properties}

Here we show that under mild nonparametric conditions, the proposed estimator is root-n consistent and asymptotically normal. \\

\begin{theorem}\label{thm:SA}
    In addition to Assumption~\ref{A1} and the $\Gamma$-selection bias condition in \eqref{eq:gammacon}, assume that 
    \begin{enumerate}
        \item $\delta\in[\delta_l,\delta_u]$ for some $0<\delta_l\leq\delta_u\leq\infty$; $\Gamma\in[\Gamma_l,\Gamma_u]$ for some $0<\Gamma_l\leq\Gamma_u\leq\infty$.
        \item $\hat{\mu}_a, \hat{\theta}_a^\pm, \hat{\nu}_a^\pm$ are bounded with probability 1, and $\hat{\nu}\geq C_l$ for some constant $C_l$ with probability 1.
        \item The conditional density $p_Y(y\mid A=a, X)$ exists, and $\sup_{x,y}p_Y(y\mid A=a, X=x) < \infty$. 
        \item $\parallel\varphi^\pm(Z;\hat{\eta},\delta)-\varphi^\pm(Z;\eta,\delta)\parallel=o_\PP(1)$. %\sup_\delta
        \item $||\Delta\pi||^2 + \sum_a\left\{||\Delta\pi||\cdot||\Delta\mu_a|| + \sum_\pm\left(||\Delta\pi||\cdot||\Delta\theta_a^\pm|| + ||\Delta\theta_a^\pm||\cdot||\Delta\nu_a^\pm|| + ||\Delta\theta_a^\pm||^2\right)\right\} = o_\PP(1/\sqrt{n})$, where $\Delta s = \hat{s} - s$ for any function $s$.
        % $||\hat{\eta}-\eta ||_2 = o_\PP(n^{-1/4})$, where $\eta = (\pi,\mu_1,\mu_0,\theta^+_1,\theta^+_0, \theta^-_1,\theta^-_0, \nu^+_1,\nu^+_0, \nu^-_1,\nu^-_0)$.
        %\item $R_2=o_\PP(1/\sqrt{n})$, where
    %\begin{equation}
    %\begin{aligned}
    %    R_2=||&\hat{\omega}-\omega||\Big( ||\hat{\pi}-\pi||+||\hat{\mu}_1-\mu_1|| + ||\hat{\mu}_0-\mu_0||
    %    + ||\hat{\mu}_0-\mu_0|| + ||\hat{\theta}^+_1-\theta^+_1|| + ||\hat{\theta}^+_0-\theta^+_0|| \\
    %    &+||\hat{\theta}^-_1-\theta^-_1|| + ||\hat{\theta}^-_0-\theta^-_0||
    %    + ||\hat{\nu}^+_1-\nu^+_1|| + ||\hat{\nu}^+_0-\nu^+_0|| +
    %    ||\hat{\nu}^-_1-\nu^-_1|| + ||\hat{\nu}^-_0-\nu^-_0||
    %    \Big)\\
    %     +& ||\hat{\pi}-\pi||\Big(||\hat{\mu}_1-\mu_1|| + ||\hat{\mu}_0-\mu_0|| + ||\hat{\theta}^+_1-\theta^+_1|| + ||\hat{\theta}^+_0-\theta^+_0|| +
    %    ||\hat{\theta}^-_1-\theta^-_1|| + ||\hat{\theta}^-_0-\theta^-_0||\\
    %    & + ||\hat{\nu}^+_1-\nu^+_1|| + ||\hat{\nu}^+_0-\nu^+_0|| +
    %    ||\hat{\nu}^-_1-\nu^-_1|| + ||\hat{\nu}^-_0-\nu^-_0||
    %    \Big) \\
    %    + &||\hat{\pi}-\pi||^2.
    %\end{aligned}
    %\end{equation}
    \end{enumerate}
    
    Then the estimator in Algorithm~\ref{algo} satisfies
\begin{equation*}
    \frac{\widehat{\psi}^\pm(\delta)-\psi^\pm(\delta)}{{\sigma}(\delta)/\sqrt{n}}\leadsto N(0,1),
\end{equation*}
where ${\sigma}^2(\delta)$ is the variance of the influence function $\varphi^\pm(Z;\eta,\delta)$.
\end{theorem}

\bigskip

Most of the conditions are standard, except perhaps Condition 3, which requires the conditional distribution of $Y$ to have bounded density. This boundedness is the key to compute and control the derivative of $g(\theta) := \E\left[f_\theta(Y)\mid A=a,X\right]$. \\

The first condition indicates that the result is valid for bounded $\delta$ and $\Gamma$. We also assume that estimated nuisance functions are bounded, that their convergence rates satisfy $||\Delta\pi||^2 + \sum_a\left\{||\Delta\pi||\cdot||\Delta\mu_a|| + \sum_\pm\left(||\Delta\pi||\cdot||\Delta\theta_a^\pm|| + ||\Delta\theta_a^\pm||\cdot||\Delta\nu_a^\pm|| + ||\Delta\theta_a^\pm||^2\right)\right\} = o_\PP(1/\sqrt{n})$, and that $\varphi^\pm$ is consistent with these estimated nuisance functions. These convergence rate conditions could be achieved if $\pi,\theta_a^\pm$ are estimated at rate $n^{-1/4}$, and $||\hat{\pi}-\pi||\cdot||\hat{\mu}_a-\mu_a||, ||\hat{\pi}-\pi||\cdot||\hat{\theta}_a^\pm-\theta_a^\pm||, ||\hat{\theta}_a^\pm-\theta_a^\pm||\cdot||\hat{\nu}_a^\pm-\nu_a^\pm||=o_\PP(n^{-1/2})$. These are standard conditions in semiparametrics and double machine learning \citep{van2003unified,  chernozhukov2018double, kennedy2024semiparametric}. \\

\section{Time-varying sensitivity model}\label{sec:longitudinal}

In this section we extend our analysis to time‐varying treatments. We discuss incremental propensity score interventions in longitudinal settings and adopt the marginal sensitivity model to relax sequential exchangeability. Under this framework, we provide equivalent characterizations of sharp bounds for incremental effects and establish their identification from observed data, while highlighting the technical challenges involved in constructing practical estimators. \\

\begin{remark}
For simplicity we consider the two time point case with $T=2$. All results can be similarly extended beyond this setting to $T > 2$.
\end{remark}

\bigskip

Here we suppose  we observe $n$ independent and identically distributed samples of $(X_1, A_1, X_2, A_2, Y)$, where $X_t$ denotes covariates at time $t$, $A_t$ denotes the binary outcome $\{0,1\}$ at time $t$, and $Y$ denotes the outcome. We use an overline for the history of a variable, and we drop the subscript when $k=T$; for example, $\Xbar = \Xbar_2=(X_1, X_2)$, $\Abar=\Abar_2=(A_1, A_2)$ for $T=2$. Similarly, we define $\abar=(a_1, a_2)$ as a specific treatment sequence with $a_1$ at time 1 and $a_2$ at time 2. We define the history up to time $k$ as $H_k=(\Xbar_k, \Abar_{k-1})$  and define  $\Abar_0$ to be null. \\

The propensity score under the incremental intervention is defined as 
\begin{align*}
    q_k(h_k; \delta, \pi_k) = \frac{\delta \pi_k(h_k)}{\delta \pi_k(h_k) + 1 - \pi_k(h_k)}.
\end{align*}
We can express the probability of each treatment strategy $\abar$ under the incremental intervention
$q(\abar, H_2; \delta) = \PP(Q(\delta)=\abar \mid H_2)$ 
as a functional of the $\pi_k$, since
\begin{align*}
    \mathrm{d} Q_k(a_k \mid h_k) = \frac{a_k \delta \pi_k(h_k) + (1-a_k) (1 - \pi_k(h_k))}{\delta \pi_k(h_k) + (1 - \pi_k(h_k))}
\end{align*}
and each $Q_k$ is randomly drawn from $Q_k(h_k; \delta, \pi_k)$ at each stage. The extensions of Assumptions \ref{A1}, \ref{A2}, \ref{A3} to longitudinal data are as follows. \\

\begin{assumption}{A}{1*}\label{A1*}{\normalfont (\textbf{Consistency})}
$
Y=Y^{\abar} \text{ if } \Abar = \abar.
$
\end{assumption}

\begin{assumption}{A}{2*}\label{A2*}{\normalfont (\textbf{No unmeasured confounding})}
$A_k \perp\!\!\!\perp \Yabar \mid (\Abar_{k-1}, \Xbar_k).
$ for $k=1,...,T.$
\end{assumption}

\begin{assumption}{A}{3*}\label{A3*}{\normalfont (\textbf{Positivity})}
$
P(A_k=a_k \mid \Abar_{k-1}=\abar_{k-1}, \Xbar_k) > 0.
$ for $k=1,...,T.$
\end{assumption}

\bigskip

As before, we do not rely on the positivity assumption due to our focus on incremental effects. To relax Assumption~\ref{A2*}, we use the marginal sensitivity model \citep{tan2006distributional}, which is related to but slightly different from Rosenbaum's model in \eqref{eq:gammacon}. \\

\subsection{Marginal sensitivity model}

The marginal sensitivity model from \citet{tan2025sensitivity} can be stated as
\begin{align}\label{eq:lambdaBD}
    \frac{1}{\Lambda_2} \leq \lambda_{2,*}^{\abar}(\Xbar, y) \leq \Lambda_2, \ \ \ 
    \frac{1}{\Lambda_1} \leq \lambda_{1,*}^{\abar}(X_1, y) \leq \Lambda_1
\end{align}
where
\begin{align}
    \lambda_{2,*}^{\abar}(\Xbar, y) &= \frac{\mathrm{d}\PP_{\Yabar}(y \mid A_1=a_1, A_2=1-a_2, \Xbar)}{\mathrm{d}\PP_{\Yabar}(y \mid A_1=a_1, A_2=a_2, \Xbar)}, \label{eq:lambda2}\\
    \lambda_{1,*}^{\abar}(X_1, y) &= \frac{\mathrm{d}\PP_{\Yabar}(y \mid A_1=1-a_1, X_1)}{\mathrm{d}\PP_{\Yabar}(y \mid A_1=a_1, X_1)} \label{eq:lambda1},
\end{align}
for $\abar=(0,0), (0,1), (1,0), (1,1)$. 
Bayes' rule shows that $\lambda^{\abar}_{k,*}$ can be expressed as odds ratios,
\begin{align*}
   \lambda_{2,*}^{\abar}(\Xbar, y) &= \frac{\PP(A_2=a_2 \mid A_1=a_1, \Xbar) / \PP(A_2=1-a_2 \mid A_1=a_1, \Xbar)}{\PP(A_2=a_2 \mid A_1=a_1, \Xbar, Y^{\abar}) / \PP(A_2=1-a_2 \mid A_1=a_1, \Xbar, Y^{\abar})}, \\
   \lambda_{1,*}^{\abar}(X_1, y) &=
   \frac{\PP(A_1=a_1\mid X_1) / \PP(A_1=1-a_1\mid X_1)}{\PP(A_1=a_1\mid X_1, Y^{\abar}) / \PP(A_1=1-a_1\mid X_1, Y^{\abar})}.
\end{align*}
Assumption~\ref{A2*} implies that $\lambda_{2,*}^{\abar} \equiv \lambda_{1,*}^{\abar} \equiv 1$, and \eqref{eq:lambdaBD} is a common relaxation of exchangeability under unmeasured confounding. \citet{tan2025sensitivity} analyzes the average treatment effect under this sensitivity model with longitudinal data.\\

\subsection{Identification}

As shown in \citet{kennedy2019nonparametric}, incremental effects are identified  under  Assumptions~\ref{A1*} and \ref{A2*}. However, only partial identification is possible without Assumption~\ref{A2*}. In this subsection, we introduce sharp bounds on the incremental effect in such a setting. \\

The following proposition shows that the incremental effect can be expressed by the expectation of potential outcomes with regard to the targeted treatment distribution. \\

\begin{proposition}\label{prop:incmt2}
    Under Assumption~\ref{A1*} , the incremental effect is given by
    \begin{align*}
        \psi(\delta) = \sum_{\abar\in \mathcal{A}^2} \E\left[Y^{\abar} q(\abar, H_2;\delta)\right].
    \end{align*}
\end{proposition}

\bigskip

By definition of incremental interventions, the condition $H_2=(X_1, A_1, X_2)$ is necessary; neither $\PP(Q(\delta) \mid X_1, A_1)$ nor $\PP(Q(\delta) \mid X_1, X_2)$ is identifiable under such interventions. \\

\begin{remark}
For the remainder of this paper, we focus on the upper bound of $\psi(\delta)$. Corresponding results for the lower bound can be derived similarly. \\
\end{remark}

There is an equivalent characterization of density ratios in \eqref{eq:lambda2} and \eqref{eq:lambda1}. \\

\begin{proposition}\label{prop:necsuff}
    Suppose that Assumption~\ref{A1*} holds for $\abar \in \mathcal{A}^2$. Then
    \begin{align}
        \E(\lambda_{2,*}^{\abar}(\Xbar, y) \mid \Abar=\abar, \Xbar) &= 1, \label{eq:compat2}\\
        \E(\lambda_{1,*}^{\abar}(X_1, y) \mid A_1=a_1, X_1) &= 1, \label{eq:compat11} \\
        \E\{\E\left(\lambda_{1,*}^{\abar}(X_1, y) \rho(\Xbar, Y;\lambda_{2,*}^{\abar}) \mid \Abar=\abar, \Xbar\right) \mid A_1=a_1, X_1)\} &= 1 \label{eq:compat1}
    \end{align}
    where 
    \begin{align}\label{eq:rho}
        \rho(\Xbar, y; \lambda_2) = \pi_1(\Xbar) + (1-\pi_1(\Xbar))\lambda_2(\Xbar, y).
    \end{align}
    If for every treatment strategy $\abar$, there is a pair of non-negative functions $\lambda_2^{\abar}(\Xbar, y)$ and $\lambda_1^{\abar}(X_1, y)$ satisfying 
    \begin{align}
        \E(\lambda_2^{\abar}(\Xbar, y) &\mid \Abar=\abar, \Xbar) = 1, \label{eq:compat2R}\\
        \E\{\E\left(\lambda_1^{\abar}(X_1, y) \rho(\Xbar, Y;\lambda_{2,*}^{\abar}) \mid \Abar=\abar, \Xbar\right) &\mid A_1=a_1, X_1)\} = 1, \label{eq:compat1R}
    \end{align}
    then there is a probability distribution $\QQ$ satisfying:
    \begin{enumerate}
        \item $\QQ$ is compatible with the observed data $(\Abar, \Xbar, Y)$,
        \item $\lambda_2^{\abar}(\Xbar, y) = \lambda_{2,\QQ}^{\abar}(\Xbar, y), \lambda_1^{\abar}(X_1, y) = \lambda_{1,\QQ}^{\abar}(X_1, y)$ for all $\abar$.
    \end{enumerate}
    Here $\lambda_{k,\QQ}^{\abar}$ is defined as the density ratio by replacing $\PP$ with $\QQ$ in the true density ratio $\lambda_{k,*}^{\abar}$.
\end{proposition}

\bigskip

The proposition implies that any non-negative functions $\lambda^{\abar}_1, \lambda^{\abar}_2$ satisfying the constraints \eqref{eq:compat2R} and \eqref{eq:compat1R} are possible candidates for the true $\lambda^{\abar}_{1,*}, \lambda^{\abar}_{2,*}$. \\

Next, we identify the sharp bounds of the incremental effect.
\begin{proposition}\label{prop:maxQ}
    Suppose Assumptions~\ref{A1*} and \ref{A3*} hold, and the bounds for density ratios in \eqref{eq:lambdaBD} hold. Then the sharp upper bound of $\psi(\delta)$ can be identified as
    \begin{align*}
        \psi(\delta)^+ = \max_\QQ \sum_{\abar\in\mathcal{A}^2} \fipw(\lambda^{\abar}_{2,\QQ})
    \end{align*}
    where
    \begin{align*}
        \fipw(\lambda^{\abar}_{2}) &= \E\left[\mathbf{1}(\Abar=\abar)\left(1 + \frac{1 - \pi_2^*}{\pi_2^*}\lambda^{\abar}_2(\Xbar, Y)\right)q(\abar, H_2;\delta)Y\right].
    \end{align*}
    Here $\pi_2^* = \PP(A_2=a_2 \mid H_2)$ and $\QQ$ can be any probability distribution that is compatible with the observed data and where \eqref{eq:lambdaBD} holds.
\end{proposition}

\bigskip

This result is an optimization with all possible distribution $\QQ$, and by Proposition~\ref{prop:necsuff}, we can provide another identification of the sharp bounds. \\

\begin{proposition}\label{prop:maxlambda}
    The sharp upper bound in Proposition~\ref{prop:maxQ} can be alternatively identified as
    \begin{align}\label{eq:maxlambda}
        \psi(\delta)^+ = \sum_{\abar\in\mathcal{A}^2} \max_{\lambda^{\abar}_1,\lambda^{\abar}_2}\fipw(\lambda^{\abar}_{2})
    \end{align}
    for any non-negative function $\lambda^{\abar}_1, \lambda^{\abar}_2$ satisfying \eqref{eq:compat2R}, \eqref{eq:compat1R} and \eqref{eq:lambdaBD}.
\end{proposition}

\bigskip

This is a quick result from Proposition~\ref{prop:necsuff}. Note that $\fipw(\lambda^{\abar}_{2})$ is dependent on $\lambda^{\abar}_2$ but not $\lambda^{\abar}_1$. However, the compatibility condition \eqref{eq:compat1R} establishes a relationship between the pair $\lambda^{\abar}_1, \lambda^{\abar}_2$. Therefore, the maximum is taken over both $\lambda^{\abar}_1$ and $\lambda^{\abar}_2$. \\

It is noteworthy that currently we only establish identification of the sharp bounds, and a practical estimator is not yet available. Additional work is required to find the dual problem of \eqref{eq:maxlambda}, where $\lambda^{\abar}_2$ appears explicitly in the objective, but $\lambda^{\abar}_1$ is constrained implicitly through iterated expectations. We leave this to future work. \\

\section{Simulation}

In this section we present simulation studies to evaluate the finite-sample performance of the proposed methods. We examine how the bounds on incremental effects vary with the sensitivity parameter and the intervention level, and assess the behavior of the doubly robust estimator under different rates of nuisance function estimation. The simulations illustrate the theoretical properties of the bounds and demonstrate the advantages of the proposed estimator relative to plug-in alternatives. \\

Suppose that the full data has the following distribution:
\begin{align*}
    X &\sim \text{Unif}(0,1), \\
    \pi(X) &= \text{expit}(X), \\
    A &\sim \text{Bernoulli}(\pi(X)), \\
    Y &\sim (1+A)X + \epsilon, \epsilon \sim N(0, 0.5^2).
\end{align*}
Therefore, $\mu_a(X) = (1+a)X.$ According to \eqref{eq:cond0}, 
\begin{align*}
    \E\left[ (Y-\theta_1^-(X))_+ - \Gamma (Y-\theta_1^-(X))_- \mid A=1, X\right] = 0.
\end{align*}
Define $h=h(\Gamma)$ as the unique solution to
\begin{align*}
    \E\left[(\xi-h)_+ -\Gamma (\xi - h)_- \right] = 0, \quad \xi\sim N(0,1).
\end{align*}
The solution is unique because the right-hand side is a non-decreasing function of $h$. Then the nuisance functions $\theta, \nu$ can be expressed as
\begin{align*}
    \theta_a^-(X) &= \mu_a(X) + \sigma \cdot h, \\
    \theta_a^+(X) &= \mu_a(X) - \sigma \cdot h, \\
    \nu_a^-(X) &= (1-\Gamma)(1-\Phi(h)) + \Gamma, \\
    \nu_a^+(X) &= (1-1/\Gamma)\Phi(h) + 1 / \Gamma
\end{align*}
where $\sigma=0.5$ and $\Phi$ is the CDF of the standard normal distribution. As we can see, the difference between $\theta_a^\pm$ and $\mu_a$ relies largely on the conditional distribution of $Y^a\mid X$, which is directly related to the distribution of $\epsilon$ in our setting. This is sensible since partial identification of the bounds is a result of unmeasured confouding, and $\theta$ attempts to capture the unobserved structure of the outcome, apart from covariates $X$. \\

\begin{figure}
    \centering
    \begin{subfigure}{0.9\textwidth}
        \centering
        \includegraphics[width=\linewidth]{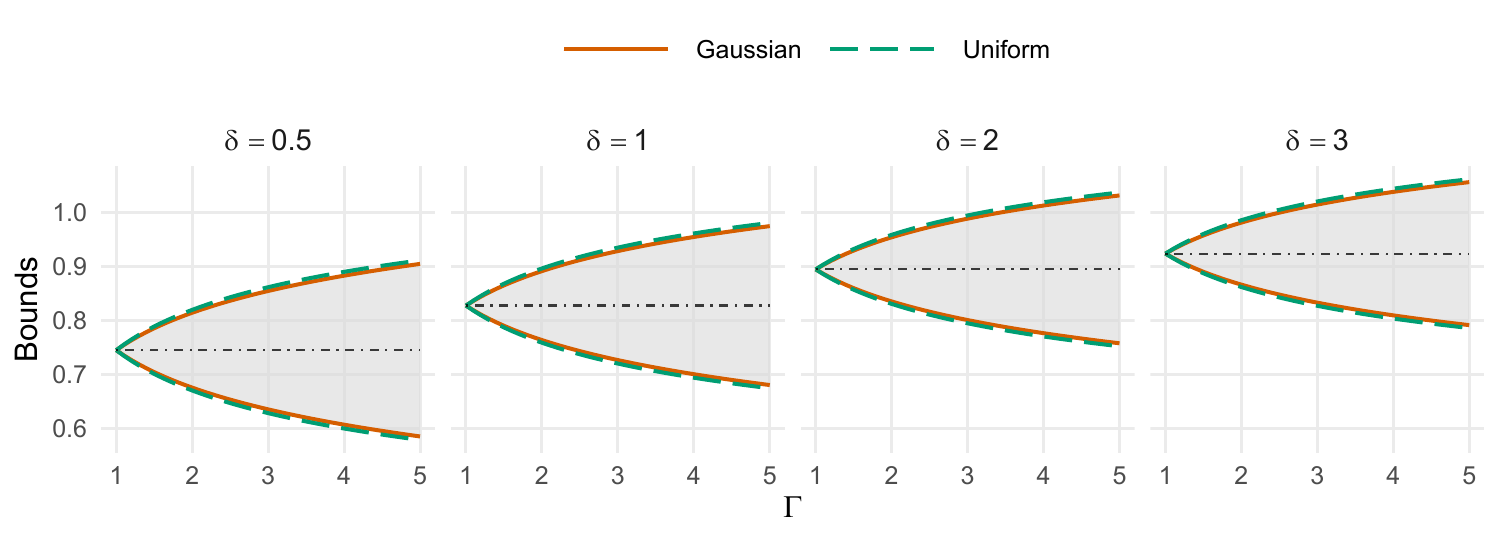}
        \caption{$X\sim\text{Unif}(0,1)$}
        \label{fig:length:a}
    \end{subfigure}
    \hfill
    \begin{subfigure}{0.9\textwidth}
        \centering
        \includegraphics[width=\linewidth]{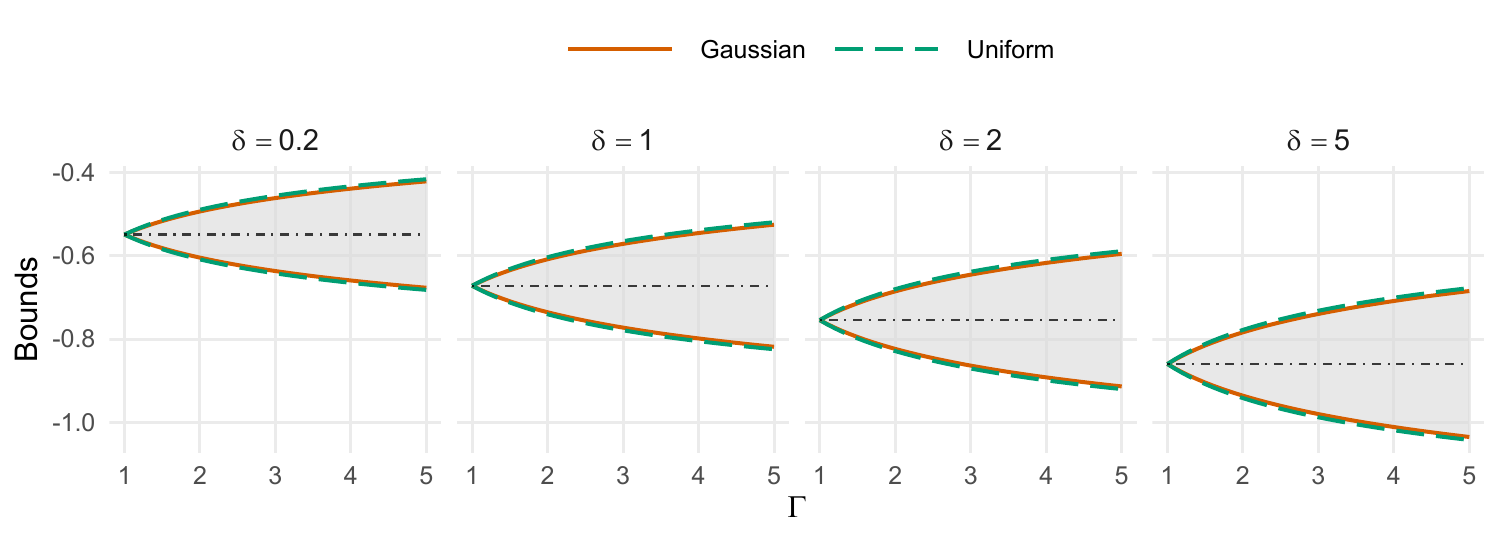}
        \caption{$X\sim\text{Unif}(-1,0)$}
        \label{fig:length:c}
    \end{subfigure}
    \hfill
    \begin{subfigure}{0.9\textwidth}
        \centering
        \includegraphics[width=\linewidth]{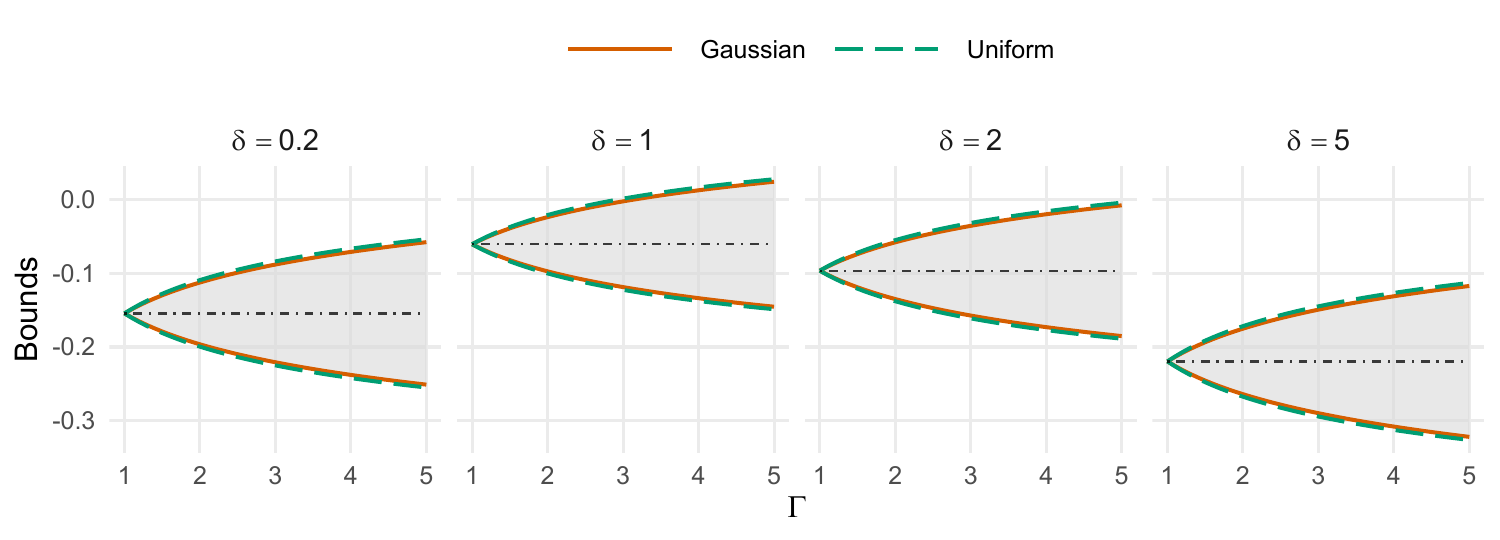}
        \caption{$X\sim\text{Unif}(-4,3)$}
        \label{fig:length:b}
    \end{subfigure}
    \caption{Plot of bounds as $\Gamma$ varies. The solid lines correspond to Gaussian noise $\epsilon\sim N(0, 0.5^2)$ in $Y$, and the dashed lines correspond to noise that is uniformly distribution with the same variance, $\text{Unif}(-0.5\sqrt{3}, 0.5\sqrt{3})$.}
    \label{fig:length}
\end{figure}

\bigskip

First, we analyze how the bounds $\psi^\pm$ vary with the confounding parameter $\Gamma$. Figure~\ref{fig:length} shows the upper and lower bounds as $\Gamma$ and $\delta$ change. The bounds are wider for larger $\Gamma$, and the exact shape of such curves depend on the conditional distribution $Y^a\mid X$. The symmetry of the lower curve and the upper curve results from the symmetric distribution of the noise in $Y$, $\epsilon$. According to Equation~\ref{eq:L'},
\begin{align*}
    \frac{\partial}{\partial\delta} L_{incremental} = -2\sigma h(\Gamma)\E\left[\frac{\pi(X)(1-\pi(X))(1-2\pi(X))}{(\delta\pi(X)+1-\pi(X))^2}\right],
\end{align*}
since $\theta_0^+(X)-\theta_0^-(X)\equiv\theta_1^+(X)-\theta_1^-(X)\equiv -2\sigma h(\Gamma)>0$ if $\Gamma>1$. We can change its sign by modifying the distribution of $X$. For example, $\frac{\partial}{\partial\delta} L_{incremental}\leq 0$ since $\pi(X)=\expit(X)\geq \frac{1}{2}$ as $X\sim \text{Unif}(0,1)$, according to the setup. \\

Figure~\ref{fig:length} shows the plot of bounds with different patterns. In Figure~\ref{fig:length:a}, the larger $\delta$ is, the narrower the bounds are; in Figure~\ref{fig:length:c}, the length is an increasing function of $\delta$; in Figure~\ref{fig:length:b}, the length decreases first and then increases. \\

Moreover, the derivative of the incremental effect, which is identified if there are no unmeasured confounders, is 
\begin{equation*}
\begin{aligned}
    \frac{\partial}{\partial\delta}\psi(\delta)
    &=\frac{\partial}{\partial\delta}\mathbb{E}\left[\frac{\delta\pi(X)\mu_1(X)+(1-\pi(X))\mu_0(X)}{\delta \pi(X)+(1-\pi(X))}\right]\\
    &=\E\left[\frac{\pi(X)(1-\pi(X))(\mu_1(X)-\mu_0(X))}{\left(\delta \pi(X)+(1-\pi(X))\right)^2}\right] \\
    &= \E\left[\frac{\pi(X)(1-\pi(X))X}{\left(\delta \pi(X)+(1-\pi(X))\right)^2}\right],
\end{aligned}
\end{equation*}
indicating there is no apparent relationship between changes in the incremental effect and changes in the length of the bounds, even under our simulation setting. \\

We also analyze the bounds when the noise in $Y$, $\epsilon$ follows the uniform distribution $\text{Unif}(-0.5\sqrt{3}, 0.5\sqrt{3})$, so that $\epsilon$ still has mean zero and variance $0.5^2$. It is also noteworthy that the bounds for uniform noise are wider than those for Gaussian noise. \\

Next, we simulate estimated nuisance functions $\hat{\eta}=(\hat{\pi}, \hat{\mu}, \hat{\theta}, \hat{\nu})$ as
\begin{align*}
    \hat{\pi}(X) &\sim \expit\left(\logit\left(\pi(X) + N(n^{-\alpha}, n^{-2\alpha})\right)\right), \\
    \hat{\mu}_a(X) &\sim \mu_a(X) + N(n^{-\alpha}, n^{-2\alpha}), \\
    \hat{\theta}_a^\pm(X) &\sim \theta_a^\pm(X) + N(n^{-\alpha}, n^{-2\alpha}), \\
    \hat{\nu}_a^\pm(X) &\sim \nu_a^\pm(X) + N(n^{-\alpha}, n^{-2\alpha}).
\end{align*}
In Theorem~\ref{thm:SA}, we proved the bias of the doubly robust estimator can be bounded by 
\begin{align*}
    ||\Delta \pi ||^2 + \sum_a\big(||\Delta\pi||\cdot||\Delta\mu_a||+\sum_\pm||\Delta\pi||\cdot||\Delta\theta_a^\pm||\big) + \sum_a\sum_\pm \big(||\Delta\theta_a^\pm||\cdot||\Delta\nu_a^\pm|| + ||\Delta\theta_a^\pm||^2 \big),
\end{align*}
and by varying the estimation rate $\alpha$, we compute the bias for different estimators. \\

\begin{figure}[h]
    \centering    
    \includegraphics[width=0.8\textwidth]{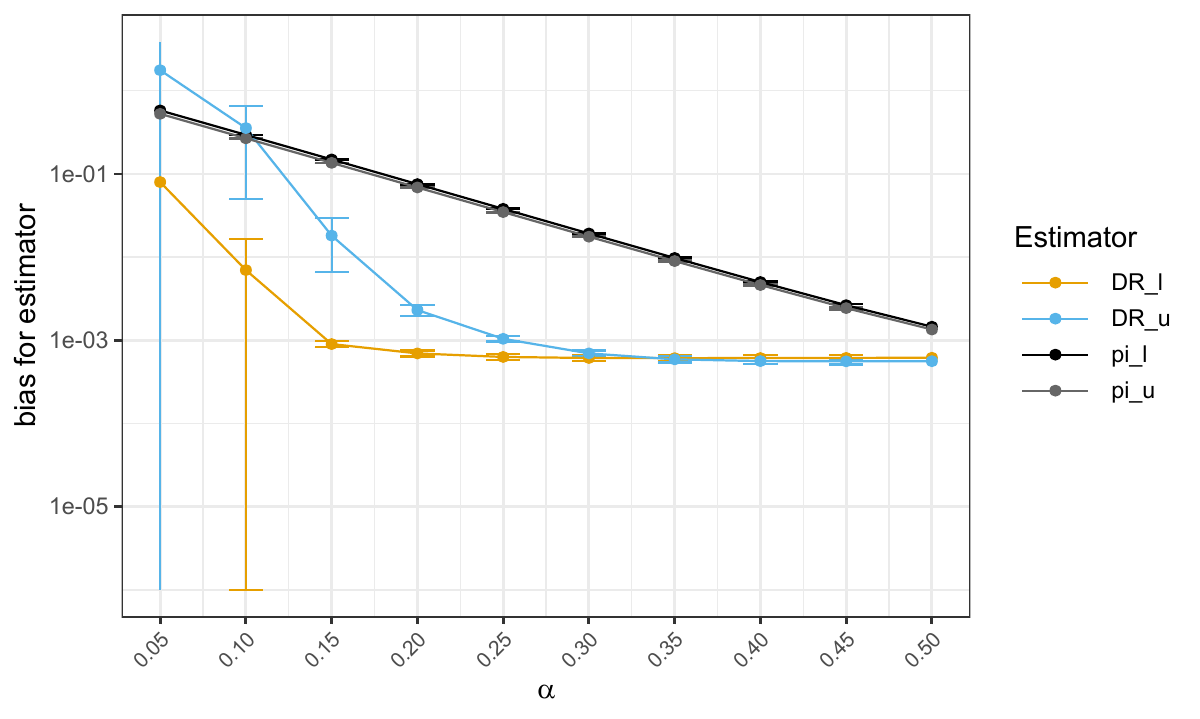}
    \caption{The absolute bias of different estimators as the estimation rate $\alpha$ varies. }
    \label{fig:bias}
\end{figure}

\bigskip

\cref{fig:bias} compares the performance of the plug-in estimator (pi) and the doubly-robust estimator (DR) based on the efficient influence function for $\delta=\Gamma=2$. The subscript $\_l$ is for the lower bound, and $\_u$ is for the upper bound. We generate 1000 samples, and the sample size is 1000. The doubly-robust estimator generally has lower bias compared with the plug-in estimator. As shown in the figure, the log-transformed bias exhibits an approximately linear relationship with $\alpha$, while the bias of doubly-robust estimators decreases at a faster rate and levels off around $\alpha=0.2$. This figure is consistent of our theory, which implies that the bias of doubly-robust estimators is upper bounded by $n^{-2\alpha}$, and the upper bound of bias for plug-in estimators is $n^{-\alpha}$. \\

\section{Data analysis}

In this section we apply the proposed sensitivity analysis framework to study the effect of victimization on subsequent offending using data from the Add Health study. We estimate incremental effects under varying intervention intensities and conduct sensitivity analyses to assess robustness to unmeasured confounding. The results illustrate how incremental interventions provide nuanced insight into the victimization–offending relationship beyond traditional average treatment effect analyses. \\

%Violence is often understood to follow a cyclical pattern. 
The cycle of violence theory posits that exposure to victimization in childhood increases the risk of delinquency and violent criminal behavior later in life \citep{widom1989cycle}. This perspective has important implications for the design of public policies aimed at mitigating long-term harms and reducing subsequent criminal behavior \citep{chen2009link}. Empirical evidence further suggests that the magnitude of these effects may vary across developmental stages, particularly during adolescence \citep{mersky2012unsafe}. Accordingly, it is important to identify and rigorously evaluate the causal impact of victimization on subsequent offending behaviors among adolescents. Related work by \citet{dulce2025effects} examines this question using the average treatment effect on the treated (ATT); in contrast, our focus on incremental effects allows us to assess robustness across a continuum of intervention intensities. \\

\subsection{Data}

We use data from the National Longitudinal Study of Adolescent Health (AddHealth)\footnote{https://addhealth.cpc.unc.edu/}, an on-going longitudinal study of a nationwide sample of adolescents in grades 7-12 in the United States from 1994-95. Information was gathered from in-school surveys and in-home interviews at five time points, which are 1995 (Wave 1), 1996 (Wave 2), 2001–02 (Wave 3), 2008–09 (Wave 4) and 2016–18 (Wave 5) \citep{harris2019cohort}.  \\

The data on each respondent can be represented as
$$
\bmZ=(X, R, RA, RY),
$$
where $X$ denotes 17 covariates   measured at Wave 1 (including demographics, exposure to violence and offending measures, etc.), $R$ is an indicator of complete observations with both the exposure and the outcome, $A$ is the exposure indicator of victimization measured at Wave 2, and $Y$ is the outcome of offending behaviors measured at Wave 3. Victimization $A$ is defined as being beaten, stabbed, shot or pulled a knife or gun at, in the year prior to the survey. Offending $Y$ is defined as the behavior of shooting, stabbing, or pointing a knife or gun at someone; robbery; burglary or stealing items valued at over \$50, in the year prior to the survey. \\

There are 6504 observations at Wave 1; out of 4834 observations at Wave 2, 4805 have observed the exposure variable $A$,  and out of 4882 observations at Wave 3, 4818 have observed the outcome variable $Y$. 3781 observations have both  the exposure and the outcome observed. For simplicity, we treat those with missing exposure but not outcome as missing both. Apart from dropout, the covariates $X$ also have a small amount of missingness (no more than 3\%). For this, missing indicators are introduced for each covariate, and included in all analyses. \\

\subsection{Analysis}

\begin{figure}[h]
    \centering    \includegraphics[width=0.75\textwidth]{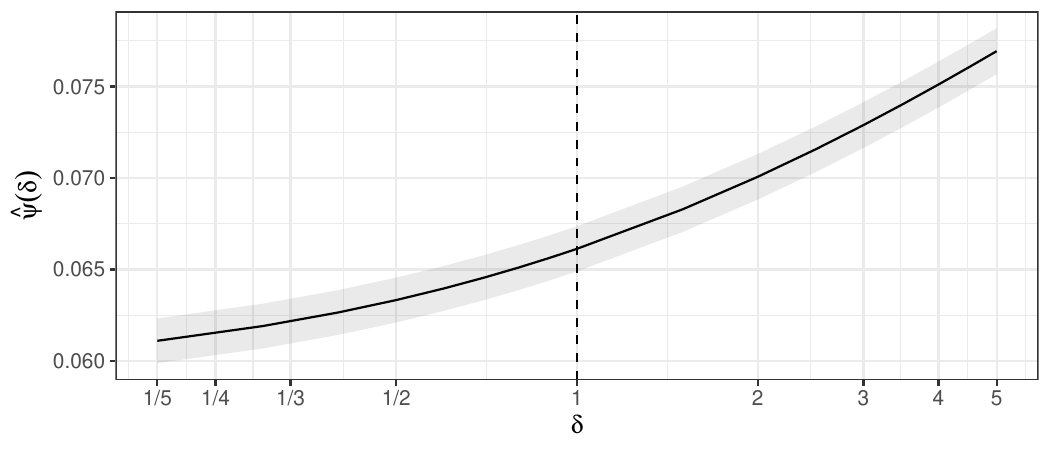}
    \caption{Estimated chance of offending if victimization odds were multiplied by $\delta$, assuming no unmeasured confounding.}
    \label{fig:delta}
\end{figure}

We use incremental interventions with $T=1$ to study the effect. We first analyze the effect of prior victimization on future offending assuming no unmeasured confounding. The causal effect under incremental interventions is estimated in Figure~\ref{fig:delta}, where the $x$-axis is $\log$ transformed. The curve appears to be convex, suggesting that the effect change is larger when the odds ratio of victimization increases, compared to when the odds ratio decreases by the same factor. Therefore, for purposes of reducing offending risk, it may be more crucial to avoid increases in the exposure of victimization, as opposed to actively reducing. \\

\begin{figure}[h]
    \centering    \includegraphics[width=0.8\textwidth]{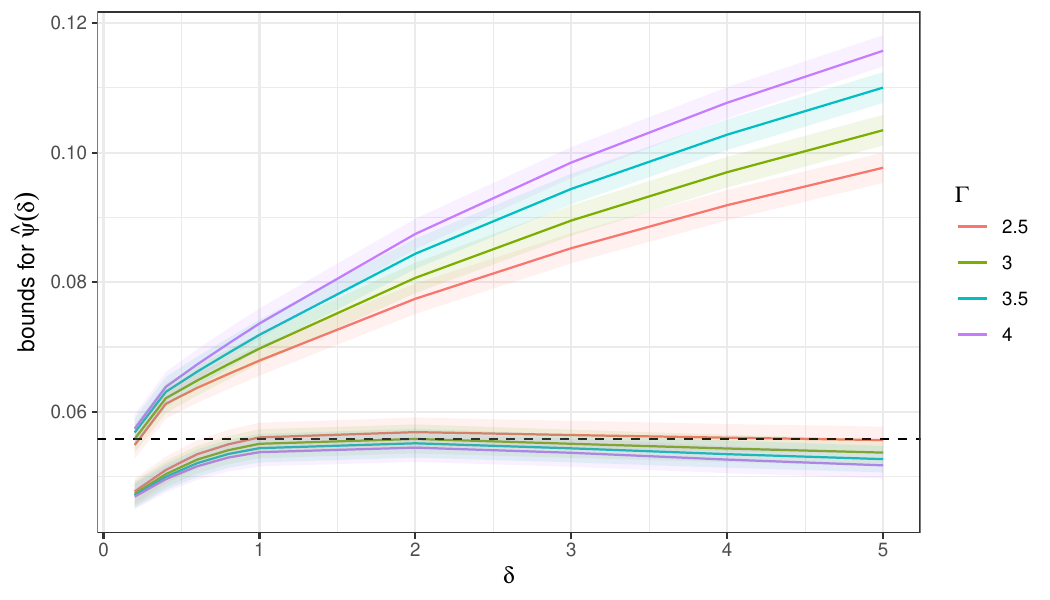}
    \caption{Plot of effect bounds as $\delta$ varies. The dashed line shows a possible effect value that remains the same under different $\delta$, for $\Gamma=3$.}
    \label{fig:bdsdelta}
\end{figure}

Next, we perform our proposed sensitivity analysis. Our analysis shows that the effect of victimization exists for $\Gamma<3$ (2.1 at the 95\% confidence level), which is relatively robust. Figure~\ref{fig:bdsdelta} shows that there lies a horizontal line between the bounds when $\Gamma\geq3$, suggesting that the effect may remain unchanged under different interventions. \\

\begin{figure}[h]
    \centering
    \includegraphics[width=0.8\textwidth]{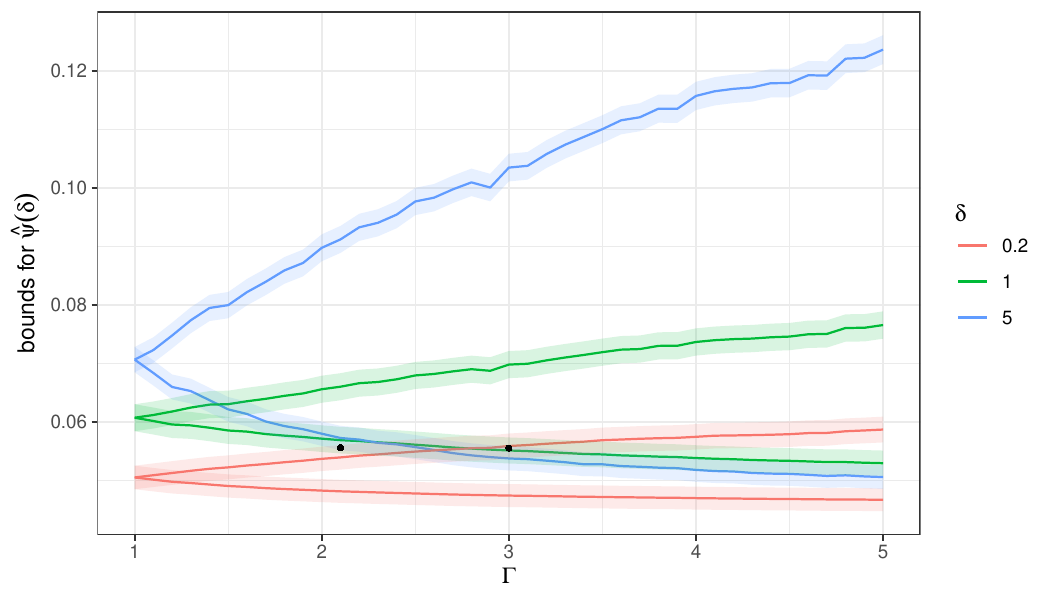}
    \caption{Plot of effect bounds as the confounding level $\Gamma$ increases. The black point with $\Gamma=3$ shows a possible value in all effect bounds as $\delta$ varies. The other point with $\Gamma=2.1$ is in all 95\% confidence intervals.}
    \label{fig:bdsgamma}
\end{figure}

Figure~\ref{fig:bdsgamma} shows how the bounds change as $\Gamma$ increases. The black point $(3, 0.0554)$ lies between the upper and lower bounds for all $\delta$ (the other black point $(2.1, 0.0555)$ is for the 95\% confidence interval). This implies that when $\Gamma=3$, the offending risk could be 5.54\%, regardless of the value of incremental intensity $\delta$. \\

Further analysis of heterogeneous incremental effects using this data set is presented in Appendix~\ref{app:heter}. \\

\section{Discussion}\label{sec:sim}

In this paper, we study sensitivity analysis for incremental effects in the presence of unmeasured confounding. For single time point settings, we adopt Rosenbaum’s sensitivity model and develop a doubly robust estimator based on the efficient influence function. Under mild conditions on the estimation of nuisance functions, the proposed estimator attains asymptotic normality. For settings with multiple time points, we work within the marginal sensitivity model and provide two equivalent characterizations of the sharp bounds for incremental effects. While these bounds are identifiable from observed data, construction of a practical estimator presents technical challenges that we leave for future work. In addition to the theoretical developments, we conduct simulation studies and an empirical analysis of the effect of victimization on subsequent offending behavior. \\

In the single time point case ($T=1$) our approach builds on bounds for $\E[Y^a \mid X]$ developed by \citet{yadlowsky2022bounds}, which are not necessarily sharp. Because the bounds on incremental effects involve combinations of multiple such components, assessing or establishing their optimality is nontrivial. An important direction for future research is therefore to derive sharp bounds for incremental effects directly under Rosenbaum’s sensitivity model. \\

For the longitudinal case ($T\geq2$), sharp bounds are identified by exploiting conditional independence of $Y^{\abar}$ given observed history $H_T$. However, a corresponding estimator has not yet been developed, as the dual optimization problem associated with Proposition~\ref{prop:maxlambda} remains unresolved. The technical difficulty arises from the fact that only $\lambda_2$ appears explicitly in the derived result, and not $\lambda_1$, complicating the task of transforming the compatibility and bound condition of $\lambda_1$ to a dual problem. Beyond resolving this issue, future work may also consider sharp bounds for conditional incremental effects of the form $\E[Y^{\abar} \mid H_t]$. Such analyses may require stronger sensitivity assumptions, as in \citet{zhang2025distributionally}. \\

\section*{Acknowledgements}

This research was supported by NSF CAREER Award 2047444. \\

\section*{References}
\vspace{-1cm}
\bibliographystyle{abbrvnat}
%\bibliography{/Volumes/flashdrive/research/bibliography}

\bibliography{proposal-references}
\setlength{\parindent}{0cm}
\appendix

\section{Appendix}

\subsection{Proof of Theorem~\ref{thm:SA}}
\begin{proof}
We only prove the theorem for the lower bound $\psi^-(\delta)$. The same reasoning can be applied for the upper bound $\psi^+(\delta)$. Note that the identification of the lower bound can be written as the summation of four terms, 
\begin{equation*}
\begin{aligned}
\psi^-(\delta) &= T_1 + T_2 + T_3 + T_4
\end{aligned}
\end{equation*}

where
\begin{align*}
    T_1 &= \E\Big[q(X)\pi(X)\mu_1(X)\Big], \\
    T_2 &= \E\Big[q(X)(1-\pi(X))\theta_1^-(X)\Big], \\
    T_3 &= \E\Big[(1-q(X))(1-\pi(X))\mu_0(X)\Big], \\
    T_4 &= \E\Big[(1-q(X))\pi(X)\theta_0^-(X)\Big],
\end{align*}
and their corresponding efficient influence functions are
\begin{align*}
    \varphi_1 &= \frac{\delta\pi(X)AY}{\delta\pi(X)+1-\pi(X)}+\frac{\delta(A-\pi(X))\pi(X)\mu_1(X)}{(\delta\pi(X)+1-\pi(X))^2}, \\
    \varphi_2 &= \frac{\delta\pi(X)(1-A)\theta_1^-(X)}{\delta\pi(X)+1-\pi(X)} + \frac{\delta(1-\pi(X))A}{\delta\pi(X)+1-\pi(X)}\frac{f_{\theta_1^-(X)}(Y)}{\nu_1^-(X)} + \frac{\delta(A-\pi(X))(1-\pi(X))\theta_1^-(X)}{(\delta\pi(X)+1-\pi(X))^2}, \\
    \varphi_3 &= \frac{(1-\pi(X))(1-A)Y}{\delta\pi(X)+1-\pi(X)} - \frac{\delta(A-\pi(X))(1-\pi(X))\mu_0(X)}{(\delta\pi(X)+1-\pi(X))^2}, \\
    \varphi_4 &= \frac{(1-\pi(X))A\theta_0^-(X)}{\delta\pi(X)+1-\pi(X)} + \frac{\pi(X)(1-A)}{\delta\pi(X)+1-\pi(X)}\frac{f_{\theta_0^-(X)}(Y)}{\nu_0^-(X)} - \frac{\delta(A-\pi(X))\pi(X)\theta_0^-(X)}{(\delta\pi(X)+1-\pi(X))^2}.
\end{align*}

Define $\hat{\varphi}_j$ as $\varphi_j$ by replacing $\eta=(\pi, \mu_a, \theta_a^\pm, \nu_a^\pm)$ with $\hat{\eta}=(\hat{\pi}, \hat{\mu}_a, \hat{\theta}_a^\pm, \hat{\nu}_a^\pm)$ in the definition. For simplicity of notation, we omit $X$ in the nuisance functions. We can show that
\begin{align*}
    \E\left(\hat{\varphi}_1 - \varphi_1\right) &= 
    \E\left[\frac{\delta(\hat{\pi}-\pi)}{\delta\pi+1-\pi}\left\{\frac{(\delta-1)(\hat{\pi}-\pi)\pi\mu_1}{(\delta\pi+1-\pi)(\delta\hat{\pi}+1-\hat{\pi})} + \frac{-\pi(\hat{\mu}_1-\mu_1)-\hat{\mu}_1(\hat{\pi}-\pi)}{\delta\hat{\pi}+1-\hat{\pi}}\right\}\right] \\
    &\lesssim \E\left[ (\hat{\pi}-\pi)^2 + |\hat{\pi}-\pi|\cdot|\hat{\mu}_1-\mu_1|\right] ,\\
    \E\left(\hat{\varphi}_2 - \varphi_2\right) &= \E\left[\frac{\delta(1-\hat{\pi})\pi}{\delta\hat{\pi}+1-\hat{\pi}}\left\{\left(1-\frac{\nu_1^-}{\hat{\nu}_1^-}\right)(\hat{\theta}_1^--\theta_1^-) + \frac{1}{2}g_1''(\theta_1^-)(\hat{\theta}_1^--\theta_1^-)^2 + o((\hat{\theta}_1^--\theta_1^-)^2)\right\} + \right.\\
    &\left.\frac{\delta(\hat{\pi}-\pi)(\hat{\theta}_1^--\theta_1^-)}{\delta\hat{\pi}+1-\hat{\pi}} + \frac{\delta(\hat{\pi}-\pi)}{\delta\hat{\pi}+1-\hat{\pi}}\left\{\frac{(\delta-1)(1-\pi)\theta_1^-(\hat{\pi}-\pi)}{(\delta\pi+1-\pi)(\delta\hat{\pi}+1-\hat{\pi})} + \frac{(\pi-1)(\hat{\theta}_1^--\theta_1^-) + \hat{\theta}_1^-(\hat{\pi}-\pi)}{\delta\hat{\pi}+1-\hat{\pi}}\right\}\right]\\
    &\lesssim \E\left[|\hat{\nu}_1^--\nu_1^-|\cdot|\hat{\theta}_1^--\theta_1^-| + (\hat{\theta}_1^--\theta_1^-)^2 + o((\hat{\theta}_1^--\theta_1^-)^2) + |\hat{\pi}-\pi|\cdot|\hat{\theta}_1^--\theta_1^-| + (\hat{\pi}-\pi)^2 \right],\\
    \E\left(\hat{\varphi}_3 - \varphi_3\right) &=  
    \E\left[\frac{\delta(\hat{\pi}-\pi)}{\delta\hat{\pi}+1-\hat{\pi}}\left\{ \frac{(\delta-1)(1-\hat{\pi})\hat{\mu}_0(\hat{\pi}-\pi)}{(\delta\hat{\pi}+1-\hat{\pi})(\delta\pi+1-\pi)} + \frac{(1-\pi)(\hat{\mu}_0-\hat{\mu})-\hat{\mu}_0(\hat{\pi}-\pi)}{\delta\pi+1-\pi}\right\} \right]\\
    &\lesssim \E\left[(\hat{\pi}-\pi)^2 + |\hat{\pi}-\pi|\cdot|\hat{\mu}_0-\mu_0|\right],\\
    \E\left(\hat{\varphi}_4 - \varphi_4\right) &= \E\left[ \frac{(1-\pi)\hat{\pi}}{\delta\hat{\pi}+1-\hat{\pi}}\left\{\left(1-\frac{\nu_0^-}{\hat{\nu}_0^-}\right)(\hat{\theta}_0^--\theta_0^-)+\frac{1}{2}g_0''(\theta_0^-)(\hat{\theta}_0^--\theta_0^-)^2 + o((\hat{\theta}_0^--\theta_0^-)^2)\right\} \right. \\
    &\left. -\frac{(\hat{\pi}-\pi)(\hat{\theta}_0^--\theta_0^-)}{\delta\hat{\pi}+1-\hat{\pi}} + \frac{\delta(\hat{\pi}-\pi)}{\delta\hat{\pi}+1-\hat{\pi}}\left\{ \frac{(\delta-1)(\hat{\pi}-\pi)}{(\delta\hat{\pi}+1-\hat{\pi})(\delta\pi+1-\pi)} + \frac{\hat{\pi}(\hat{\theta}_0^--\theta_0^-) + \hat{\theta}_0^-(\hat{\pi}-\pi)}{\delta\pi+1-\pi}\right\} \right] \\
    &\lesssim \E\left[|\hat{\nu}_0^--\nu_0^-|\cdot|\hat{\theta}_0^--\theta_0^-| + (\hat{\theta}_0^--\theta_0^-)^2 + o((\hat{\theta}_0^--\theta_0^-)^2) + |\hat{\pi}-\pi|\cdot|\hat{\theta}_0^--\theta_0^-| + (\hat{\pi}-\pi)^2 \right],
\end{align*}
where
\begin{align*}
    g_a(\theta) = \E(f_{\theta}(Y)\mid A=a, X) =\E((Y-\theta)_+-\Gamma(Y-\theta)_-\mid A=a, X).
\end{align*}

We only prove the result for $\E\left(\hat{\varphi}_2 - \varphi_2\right)$; in the proof, let $\theta=\theta_1^-, g=g_1,\nu=\nu_1^-$ for brevity. To begin with, we calculate the derivatives of $g(\theta) = \E(f_\theta(Y)\mid A=1, X)$. By the dominated convergence theorem,
\begin{align*}
    g'(\theta) &= \frac{\partial}{\partial \theta}\E(f_\theta(Y)\mid A=1,X) \\
    & =\E\left[\frac{\partial}{\partial\theta}\left((Y-\theta)_= - \Gamma(Y-\theta)_-\right)\mid A=1, X\right] \\
    &= \E\left[-\mathbf{1}(Y>\theta)-\Gamma\mathbf{1}(Y<\theta)\mid A=1, X\right] \\
    &= -\PP(Y>\theta\mid A=1, X) - \Gamma\PP(Y<\theta\mid A=1, X)  \\
    &= -(\Gamma-1)\PP(Y<\theta\mid A=1, X)-1 \\
    &= -\nu(X), \\
    g''(\theta) &= -(\Gamma-1) p_y(\theta\mid A=1,X),
\end{align*}
where $p_y(\cdot\mid A=1,X)$ is the pdf of $Y$ conditional on $A=1,X$. Therefore, according to Taylor expansion and \eqref{eq:cond0},
\begin{align*}
    g(\hat{\theta}) &= g(\theta) +g'(\theta)(\hat{\theta}-\theta)+\frac{1}{2}g''(\theta)(\hat{\theta}-\theta)^2 + o((\hat{\theta}-\theta)^2) \\
    &= -\nu(\hat{\theta}-\theta) + \frac{1}{2}g''(\theta)(\hat{\theta}-\theta)^2 + o((\hat{\theta}-\theta)^2).
\end{align*}

The expansion is taken around $\theta$ rather than $\hat{\theta}$, since 
\begin{align*}
    g'(\theta) &=-\nu(X), \\
    g'(\hat{\theta}) &=-(\Gamma-1)\PP(Y<\hat{\theta}\mid A=1, X)-1 \neq \hat{\nu}.
\end{align*}

Therefore,
\begin{align*}
    \E\left(\hat{\varphi}_2 - \varphi_2\right) &= \E\left[\frac{\delta\hat{\pi}(1-A)\hat{\theta}}{\delta\hat{\pi}+1-\hat{\pi}} + \frac{\delta(1-\hat{\pi})A}{\delta\hat{\pi}+1-\hat{\pi}}\frac{f_{\hat{\theta}}(Y)}{\hat{\nu}}+\frac{\delta(A-\hat{\pi})(1-\hat{\pi})\hat{\theta}}{(\delta\hat{\pi}+1-\hat{\pi})^2} - \frac{\delta\pi(1-A)\theta}{\delta\pi+1-\pi}\right] \\
    &= \E\left[\textcolor{cyan}{\frac{\delta\hat{\pi}(1-\pi)\hat{\theta}}{\delta\hat{\pi}+1-\hat{\pi}}} + \frac{\delta(1-\hat{\pi})\pi}{\delta\hat{\pi}+1-\hat{\pi}}\frac{g(\hat{\theta})}{\hat{\nu}} + \textcolor{blue}{\frac{\delta(\pi-\hat{\pi})(1-\hat{\pi})\hat{\theta}}{(\delta\hat{\pi}+1-\hat{\pi})^2}} - \textcolor{red}{\frac{\delta\pi(1-\pi)\theta}{\delta\pi+1-\pi}}\right] \\
    &= \E\left[\textcolor{cyan}{\frac{\delta\hat{\pi}(1-\pi)\hat{\theta}}{\delta\hat{\pi}+1-\hat{\pi}}} +\frac{\delta(1-\hat{\pi})\pi}{\delta\hat{\pi}+1-\hat{\pi}}\frac{1}{\hat{\nu}}\left(-\nu(\hat{\theta}-\theta) + \frac{1}{2}g''(\theta)(\hat{\theta}-\theta)^2 + o((\hat{\theta}-\theta)^2)\right) \right. \\
    &\left. \qquad +\textcolor{blue}{\frac{\delta(\pi-\hat{\pi})(1-\hat{\pi})\hat{\theta}}{(\delta\hat{\pi}+1-\hat{\pi})^2}} - \textcolor{red}{\frac{\delta\pi(1-\pi)\theta}{\delta\pi+1-\pi}}\right] \\
    &= \E\left[ \textcolor{cyan}{\frac{\delta\hat{\pi}(1-\pi)\hat{\theta}}{\delta\hat{\pi}+1-\hat{\pi}}} + \frac{\delta(1-\hat{\pi})\pi}{\delta\hat{\pi}+1-\hat{\pi}}\left\{\left(1-\frac{\nu}{\hat{\nu}}\right)(\hat{\theta}-\theta) + \frac{1}{2}g''(\theta)(\hat{\theta}-\theta)^2 + o((\hat{\theta}-\theta)^2)\right\} \right. \\
    &\left. \qquad  + \left(\frac{\delta\hat{\pi}(1-\pi)}{\delta\hat{\pi}+1-\hat{\pi}} - \frac{\delta\pi(1-\hat{\pi})}{\delta\hat{\pi}+1-\hat{\pi}}\right)(\hat{\theta}-\theta) - \frac{\delta\hat{\pi}(1-\pi)}{\delta\hat{\pi}+1-\hat{\pi}}(\hat{\theta}-\theta)  \right. \\
    &\left. \qquad +\textcolor{blue}{\frac{\delta(\pi-\hat{\pi})(1-\hat{\pi})\hat{\theta}}{(\delta\hat{\pi}+1-\hat{\pi})^2}} - \textcolor{red}{\frac{\delta\pi(1-\pi)\theta}{\delta\pi+1-\pi}}\right] \\
    &= \E\left[\frac{\delta(1-\hat{\pi})\pi}{\delta\hat{\pi}+1-\hat{\pi}}\left\{\left(1-\frac{\nu}{\hat{\nu}}\right)(\hat{\theta}-\theta) + \frac{1}{2}g''(\theta)(\hat{\theta}-\theta)^2 + o((\hat{\theta}-\theta)^2)\right\} \right. \\
    &\left. \qquad + \frac{\delta(\hat{\pi}-\pi)(\hat{\theta}-\theta)}{\delta\hat{\pi}+1-\hat{\pi}} + \frac{\delta\hat{\pi}(1-\pi)\theta}{\delta\hat{\pi}+1-\hat{\pi}} +\textcolor{blue}{\frac{\delta(\pi-\hat{\pi})(1-\hat{\pi})\hat{\theta}}{(\delta\hat{\pi}+1-\hat{\pi})^2}} - \textcolor{red}{\frac{\delta\pi(1-\pi)\theta}{\delta\pi+1-\pi}}\right] \\
    &= \E\left[\frac{\delta(1-\hat{\pi})\pi}{\delta\hat{\pi}+1-\hat{\pi}}\left\{\left(1-\frac{\nu}{\hat{\nu}}\right)(\hat{\theta}-\theta) + \frac{1}{2}g''(\theta)(\hat{\theta}-\theta)^2 + o((\hat{\theta}-\theta)^2)\right\} \right. \\
    &\left. \qquad + \frac{\delta(\hat{\pi}-\pi)(\hat{\theta}-\theta)}{\delta\hat{\pi}+1-\hat{\pi}} + \textcolor{blue}{\frac{\delta(\pi-\hat{\pi})(1-\hat{\pi})\hat{\theta}}{(\delta\hat{\pi}+1-\hat{\pi})^2}} + \frac{\delta(1-\pi)(\hat{\pi}-\pi)\theta}{(\delta\hat{\pi}+1-\hat{\pi})(\delta\pi+1-\pi)}\right] \\
    &= \E\left[\frac{\delta(1-\hat{\pi})\pi}{\delta\hat{\pi}+1-\hat{\pi}}\left\{\left(1-\frac{\nu}{\hat{\nu}}\right)(\hat{\theta}-\theta) + \frac{1}{2}g''(\theta)(\hat{\theta}-\theta)^2 + o((\hat{\theta}-\theta)^2)\right\} \right. \\
    &\left. + \frac{\delta(\hat{\pi}-\pi)(\hat{\theta}-\theta)}{\delta\hat{\pi}+1-\hat{\pi}} + \frac{\delta(\hat{\pi}-\pi)}{(\delta\hat{\pi}+1-\hat{\pi}}\left\{\frac{(\delta-1)(1-\pi)\theta_1^-(\hat{\pi}-\pi)}{(\delta\pi+1-\pi)(\delta\hat{\pi}+1-\hat{\pi})} + \frac{(\pi-1)(\hat{\theta}-\theta) + \hat{\theta}(\hat{\pi}-\pi)}{\delta\hat{\pi}+1-\hat{\pi}}\right\}\right].
\end{align*}
The von Mises expansion suggests that for some estimand $\psi$ \citep{kennedy2019nonparametric},
\begin{equation*}
    \psi(\hat{\PP})-\psi(\PP) = -\int\varphi_0(z;\hat{\PP})\mathrm{d}\PP(z) + R_2(\hat{\PP},\PP).
\end{equation*}

For simplicity, $\Delta s=\hat{s}-s$ for some funtion $s$. Prior results indicate that for the estimand $\psi=\psi^-$,
\begin{align*}
    R_2(\hat{\PP},\PP) &= \sum_{j=1,2,3,4} \E\left(\hat{\varphi}_j - \varphi_j\right) \\
    &\lesssim \E\left[(\hat{\pi}-\pi)^2 + |\hat{\pi}-\pi|\cdot|\hat{\mu}_1-\mu_1| + |\hat{\nu}_1^--\nu_1^-|\cdot|\hat{\theta}_1^--\theta_1^-| + (\hat{\theta}_1^--\theta_1^-)^2 + o((\hat{\theta}_1^--\theta_1^-)^2) + |\hat{\pi}-\pi|\cdot|\hat{\theta}_1^--\theta_1^-| \right. \\
    &\left. \qquad |\hat{\pi}-\pi|\cdot|\hat{\mu}_0-\mu_0| + |\hat{\nu}_0^--\nu_0^-|\cdot|\hat{\theta}_0^--\theta_0^-| + (\hat{\theta}_0^--\theta_0^-)^2 + o((\hat{\theta}_0^--\theta_0^-)^2) + |\hat{\pi}-\pi|\cdot|\hat{\theta}_0^--\theta_0^-|\right] \\
    &\lesssim ||\Delta\pi||^2 + \sum_a\left\{||\Delta\pi||\left(||\Delta\mu_a||+||\Delta\theta_a^-||\right) + ||\Delta\theta_a^-||\cdot||\Delta\nu_a^-||+||\Delta\theta_a^-||^2\right\} 
    = o_\PP(1/\sqrt{n}).
\end{align*}

Note that 
\begin{equation*}
\begin{aligned}
    \hat{\psi}-\psi=&\psi(\hat{\PP})+\PP_n\left\{\varphi_0(Z;\hat{\PP})\right\}-\psi(\PP) \\
    = & (\PP_n-\PP)\left\{\varphi_0(Z;\PP)\right\} + (\PP_n-\PP)\left\{\varphi_0(Z;\hat{\PP})-\varphi_0(Z;\PP)\right\} + R_2(\hat{\PP}-\PP)\\
    = & (\PP_n-\PP)\left\{\varphi_0(Z;\PP)\right\} + (\PP_n-\PP)(o_\PP(1)) + o_\PP(1/\sqrt{n}).
\end{aligned}
\end{equation*}

Therefore, the central limit theorem implies that
\begin{equation}\label{eq:asy}
    \frac{\hat{\psi}-\psi}{\hat{\sigma}/\sqrt{n}}\leadsto N(0,1),
\end{equation}
which suggests asymptotical normality. Consequently, the limiting distribution of $\hat{\theta}$ in Theorem~\ref{thm:SA} is proved.
\end{proof}

\subsection{Proofs for $T=2$}
\subsubsection{Proof of Proposition~\ref{prop:necsuff}}
\begin{proof}
    It is a straightforward extension of Lemma 1 in \citet{tan2025sensitivity}, which shows a similar result for only one strategy $\abar=(1,1)$. \eqref{eq:compat2} and \eqref{eq:compat11} are trivial by definition \eqref{eq:lambda2} and \eqref{eq:lambda1}. \eqref{eq:compat1} can be shown following the same argument for $\abar=(1,1)$.

    The construction of $\QQ$ follows the same reasoning as the original proof for $\abar=(1,1).$ We extend the definition of $\mathrm{d}\QQ_{\Yabar}(y|A_1=b_1, A_2=b_2, \Xbar)$ for all $\abar$. The only noteworthy difference is that 
    \begin{align*}
        \mathrm{d}\QQ_{Y^{(0,0)},Y^{(0,1)},Y^{(1,0)},Y^{(1,1)}}(\cdot \mid \Abar, \Xbar) = \prod_{\abar\in\mathcal{A}^2}\mathrm{d} \PP_{Y^{\abar}}(\cdot\mid \Abar, \Xbar).
    \end{align*}
    We can show that the marginal distribution of $Y^{\abar}$ under $\QQ$ is compatible with $\PP$, and that $\{Y^{\abar}: \abar\in\mathcal{A}^2\}$ are conditionally independent given $\Abar, \Xbar$ under $\QQ$.
\end{proof}

\subsubsection{Proof of Proposition~\ref{prop:maxQ}}
\begin{proof}
By definition of $\lambda^*_1, \lambda^*_2$, 
    \begin{align*}
        \frac{1}{\PP(A_2=a_2 \mid A_1=a_1, \Xbar, Y^{\abar})} = 1 + \frac{1 - \pi_2^*}{\pi_2^*}\lambda^{\abar}_{2,*}(\Xbar, Y^{\abar}).
    \end{align*}

    Therefore,
    \begin{align*}
        \E\left[\mathbf{1}(\Abar=\abar)\left(1 + \frac{1 - \pi_2^*}{\pi_2^*}\lambda^{\abar}_{2,*}(\Xbar, Y)\right)q(\abar, H_2;\delta)Y\right] 
        &= \E\left[\mathbf{1}(\Abar=\abar)\left(1 + \frac{1 - \pi_2^*}{\pi_2^*}\lambda^{\abar}_{2,*}(\Xbar, Y^{\abar})\right)q(\abar, H_2;\delta)Y^{\abar}\right] \\
        %&= \E\left[\frac{\mathbf{1}(\Abar=\abar)q(\abar,H_2;\delta)Y}{\PP(A_2=a_2 \mid A_1=a_1, \Xbar, Y)}\right] \\
        &= \E\left[\frac{\mathbf{1}(\Abar=\abar)q(\abar,H_2;\delta)Y^{\abar}}{\PP(A_2=a_2 \mid A_1=a_1, \Xbar, Y^{\abar})}\right] \\
        &= \E\left[\frac{\E(\mathbf{1}(\Abar=\abar) \mid H_2, Y^{\abar})q(\abar,H_2;\delta)Y^{\abar}}{\PP(A_2=a_2 \mid A_1=a_1, \Xbar, Y^{\abar})} \right] \\
        &= \E[q(\abar,H_2;\delta)Y^{\abar}].
    \end{align*}

    According to Proposition~\ref{prop:incmt2},
    \begin{align*}
        \psi(\delta) &= \sum_{\abar} \E\left[Y^{\abar} q(\abar, H_2;\delta)\right] \\
        &= \sum_{\abar} \E\left[\mathbf{1}(\Abar=\abar)\left(1 + \frac{1 - \pi_2^*}{\pi_2^*}\lambda^{\abar}_{2,*}(\Xbar, Y)\right)q(\abar, H_2;\delta)Y\right].
    \end{align*}
    This is true for any $\QQ$ compatible with the observed data, replacing $\lambda^{\abar}_{2,*}$ with $\lambda^{\abar}_{2,\QQ}$. This completes the proof of the proposition.
\end{proof}

\subsubsection{Proof of Proposition~\ref{prop:incmt2}}

We have
    \begin{align*}
        \psi(\delta) &= \E(Y^{Q(\delta)}) =\E[\E(Y\mid H_2)] \\
        &= \sum_{\abar} \E\left[Y^{\abar} \cdot \PP(Q(\delta)=\abar \mid H_2)\right] 
        = \sum_{\abar\in \mathcal{A}^2} \E\left[Y^{\abar} q(\abar, H_2;\delta)\right].
    \end{align*}

\subsection{Heterogeneous effects}\label{app:heter}
\begin{figure}[h]
    \centering
    \includegraphics[width=\textwidth]{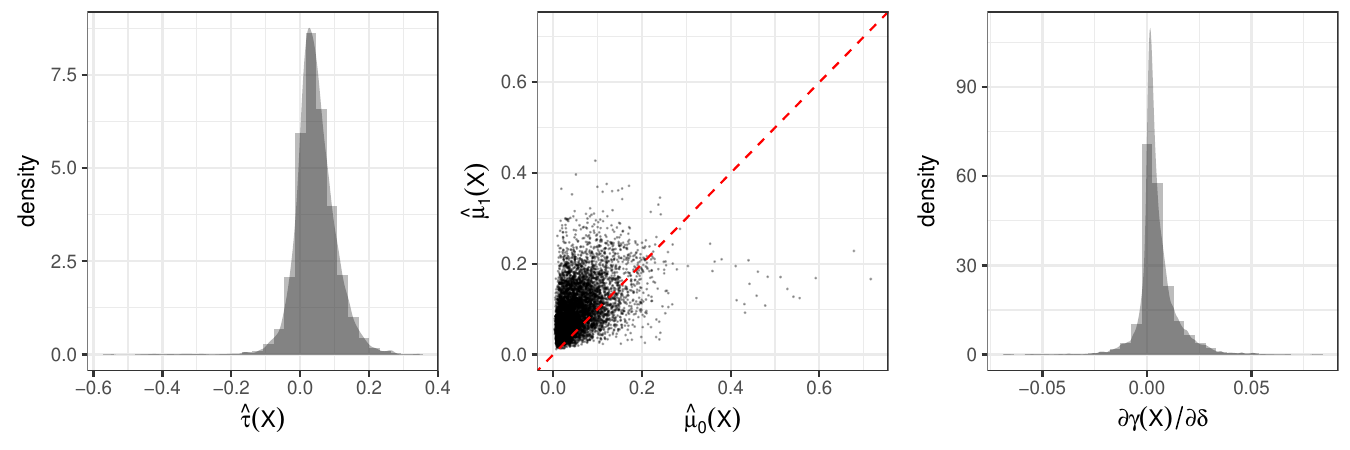}
    \caption{The plots of $\tau$, $\mu_1$ vs $\mu_0$ and $\partial\gamma/\partial\delta$, which imply that the effect is heterogeneous.}
    \label{fig:deriv}
\end{figure}

Note
\begin{equation*}
\begin{aligned}
    \frac{\partial}{\partial\delta}\psi(\delta)
    &=\frac{\partial}{\partial\delta}\mathbb{E}\left[\frac{\delta\pi(X)\mu_1(X)+(1-\pi(X))\mu_0(X)}{\delta \pi(X)+(1-\pi(X))}\right]\\
    &=\E\left[\frac{\pi(X)(1-\pi(X))(\mu_1(X)-\mu_0(X))}{\left(\delta \pi(X)+(1-\pi(X))\right)^2}\right],
\end{aligned}
\end{equation*}
which is $\pi(X)(1-\pi(X))(\mu_1(X)-\mu_0(X))$ when $\delta=1$. A plug-in estimator for the derivative, along with $\mu_1(X),\mu_0(X)$ and ATE $\tau(X)=\mu_1(X)-\mu_0(X)$, are presented in Figure~\ref{fig:deriv}. There is some spread in the distribution of both $\tau$ and $\partial\gamma/\partial\delta$, which suggests that the effect varies with $X$, i.e., there is some heterogeneity in effects. This is also shown in the comparison between $\hat{\pi}_0$ and $\hat{\pi}_1$.

However, it is unreasonable to make a direct comparison between the plot of $\tau$ and the plot of $\partial\gamma/\partial\delta$. The former is based on the interventions of $\pi(X)=1$ and $\pi(X)=0$ (every individual is exposed to victimization, or every individual is in the control group). The latter is based on the incremental propensity score intervention, which multiplies the odds of victimization by $\delta$. It is a softer intervention compared with complete exposure or complete control. The derivative is evaluated at $\delta=1$, which shows how the effect would vary if the victimization process were shifted infinitesimally from the observational level.

\begin{figure}[h]
    \centering
    \includegraphics[width=0.95\textwidth]{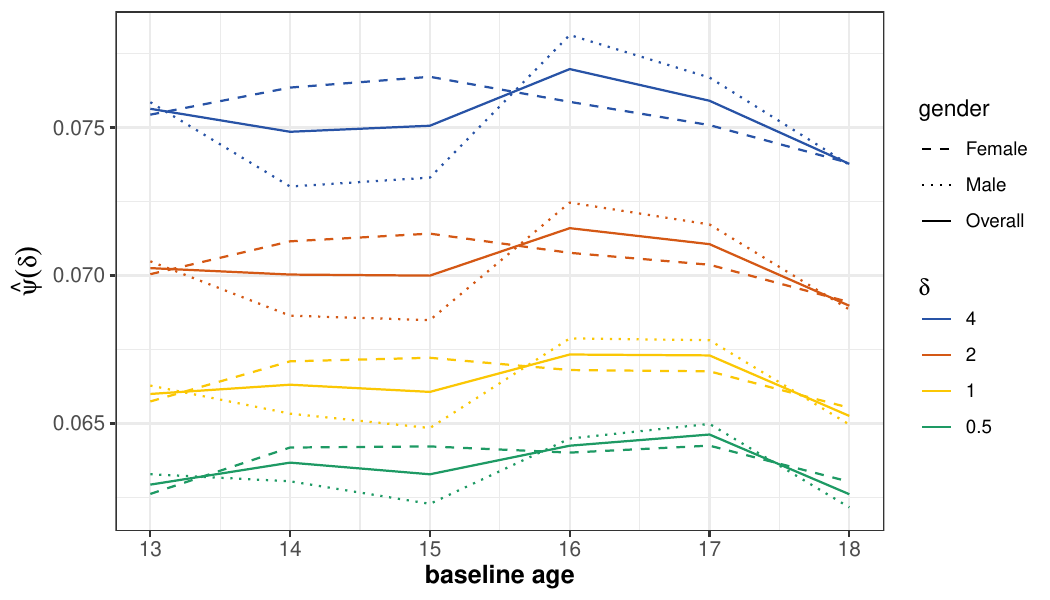}
    \caption{Heterogeneous effects for baseline age and gender.}
    \label{fig:effectFage}
\end{figure}

\begin{figure}[h]
    \centering
    \includegraphics[width=0.95\textwidth]{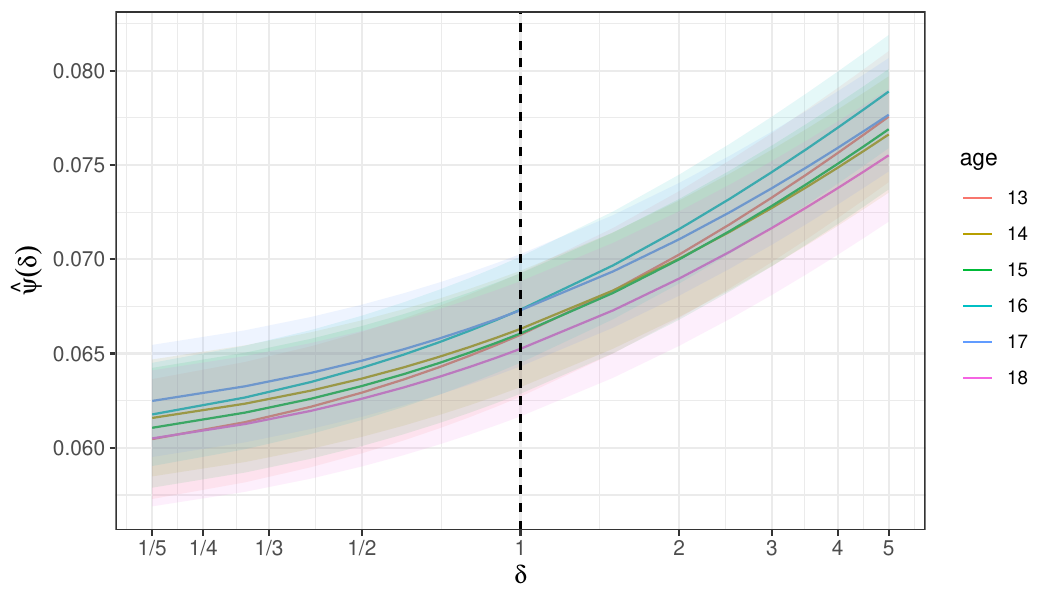}
    \caption{Confidence intervals for the heterogeneous effects grouped by baseline age.}
    \label{fig:ageCI}
\end{figure}

\citet{dulce2025effects} provide positive evidence for a heterogeneous average treatment effect on the treated (ATT) due to age. Therefore, in Figure~\ref{fig:effectFage}, we analyze how the effect changes with age and gender. In the overall population, the effect reaches a peak at the baseline age of 16-17. The estimation for male adolescents follows the same trend, while the effect for female adolescents peaks at age 15. However, as shown in Figure~\ref{fig:ageCI}, the pointwise 95\% confidence intervals for different baseline ages overlap with each other. Therefore, the heterogeneous effect of victimization is not significant across age. Similarly, we can show that gender is not a significant factor, either. 

% \begin{verbatim}
%code here
% \end{verbatim}

\end{document}